\documentclass[twocolumn,pdflatex,sn-mathphys-num]{sn-jnl}

\usepackage{booktabs}
\usepackage{graphicx}
\usepackage{subcaption}
\usepackage{float}
\usepackage{tabularx}
\usepackage{amsmath}
\usepackage{multirow,array}
\usepackage{makecell}
\usepackage{indentfirst}
\usepackage{xcolor}
\usepackage{url}
\usepackage[normalem]{ulem}
\usepackage{geometry}
\usepackage{stfloats}
\usepackage{multirow,tabularx}
\usepackage{amssymb}
\usepackage{booktabs}
\usepackage{tabularx}
\begin{document}

\title[A Priority-Aware Dual-Channel Feature Fusion]{A Priority-Aware Dual-Channel Feature Fusion Method for Urban Rail Service Traffic Classification}

\author[1]{\fnm{Xinpeng} \sur{Liu}}\email{24125047@bjtu.edu.cn}

\author*[1]{\fnm{Junhui} \sur{Zhao}}\email{junhuizhao@hotmail.com}

\author[1]{\fnm{Zhengyuan} \sur{Wu}}\email{23111011@bjtu.edu.cn}

\author[1]{\fnm{Huaicheng} \sur{Li}}\email{huaichengli@bjtu.edu.cn}

\affil*[1]{\orgdiv{School of Electronic and Information Engineering,}\orgname{Beijing Jiaotong University,}\orgaddress{\city{Beijing,}\postcode{100044,}\country{China}}}

\abstract{
With the deep integration of 5G and IoT in urban rail transit, service traffic grows explosively and accurate classification of heterogeneous service flows is essential for safe and efficient railway operations. Urban rail communication systems are subject not only to operational disturbances such as equipment failures and maintenance interference, but also to complex transmission patterns characterized by the interleaving of multi-service traffic flows. Under these operating conditions, the widespread deployment of proprietary protocols and the high prevalence of encrypted traffic further diminish the applicability of conventional port-based and Deep Packet Inspection (DPI) classification methods.
To address these challenges, we propose a priority-aware dual-channel feature fusion framework. Raw traffic bytes and statistical features are mapped into grayscale images and processed by a dual-branch architecture: a transfer learning-enhanced ResNet extracts fine-grained byte-level textures, while a lightweight CNN captures macroscopic statistical patterns. A channel attention mechanism dynamically recalibrates cross-modal features, and a novel Priority-Sensitive Loss (PSL) that integrates business-criticality awareness with class-balance weighting to maximize recall for safety-critical services. Evaluated on a real-world urban rail dataset using priority-weighted metrics, the method achieves 98.74\% accuracy and 99.24\% weighted recall, with recall of 99.94\% and 99.41\% on the two safety-critical services:Communication-Based Train Control(CBTC) and Emergency Radio Dispatch(ERD), providing a reliable classification foundation for priority-aware resource scheduling in urban rail communications. With only 0.18M parameters and 0.4--0.7 ms end-to-end latency, offering an excellent balance between high-precision classification, low inference latency, and edge-deployment feasibility.}

\keywords{Urban Rail Service, Traffic Classification, Dual-Channel Feature Fusion, Transfer Learning}

\maketitle

\section{Introduction}

In recent years, with the rapid development of China's high-speed railway and urban rail transit, the railway transportation system has become an important pillar of the national comprehensive transportation network, playing a key role in promoting the coordinated development of regional economies and improving social operation efficiency~\cite{ZHAO2024102484}. In this context, railway communication systems are evolving from traditional single-service bearer to multi-service converged bearer, with continuous growth of various service traffic and a significant increase in network load~\cite{10879075}. Specifically, modern railway communication networks need to simultaneously support diversified services such as train operation control, dispatching and command, passenger information service, high-definition video surveillance, and intelligent operation and maintenance Internet of Things(IoT)~\cite{chen2025advanced}, resulting in explosive growth and highly dynamic characteristics of data flows. Meanwhile, different services exhibit remarkable differences in delay, bandwidth, and reliability. For instance, mission-critical railway services such as train control and signaling systems require stringent latency and reliability guarantees,  while on-board video backhaul and equipment monitoring data focus more on large bandwidth and fault tolerance. Such diversification and complexity of service requirements impose higher demands on the service capability of communication systems.

Against the background of concurrent multi-services and heterogeneous Quality of Service(QoS) requirements, how to achieve accurate classification and efficient classification of service traffic in railway communication networks has become a prerequisite and key issue for network slicing, differentiated scheduling, and QoS guarantee~\cite{wu2026intelligent}.
However, the currently deployed service classification methods in railway communication systems mostly rely on port matching or fixed rules tailored to specific systems~\cite{Pareit2012}. Since different lines and different vendors often adopt proprietary or customized protocols, and with the continuous iteration of services and the emergence of abnormal traffic patterns caused by system faults, traditional methods suffer from severe deficiencies in recognition accuracy and generalization capability when facing protocol heterogeneity, dynamic service changes, and unknown fault scenarios.

Network traffic classification and railway communication service traffic classification share natural homology and task similarity. Both belong to traffic classification tasks in communication networks, are transmitted based on the TCP/IP protocol stack at the bottom layer, and are highly consistent in data structure, traffic generation mechanism, feature extraction method, and classification objective~\cite{AZAB2024676}. Specifically, both general Internet traffic and railway dedicated data flows can construct a characterization space through multi-dimensional features such as packet header information, payload distribution, packet length sequence, inter-arrival time, and flow duration. Their underlying logic relies on mining the spatial-temporal correlation and behavior patterns of traffic. Although railway services have domain-specific characteristics such as dedicated scenarios, high real-time requirements, and closed protocols (e.g., some adopt customized application-layer protocols or dedicated bearer networks), their core task is essentially identical to general network traffic classification, that is, to distinguish different service types by learning the temporal features, statistical features, and transmission patterns of traffic. 

Based on such commonality of underlying mechanisms, the mature feature engineering paradigms and deep learning architectures in general network traffic classification can provide a solid technical foundation for railway scenarios. Only through adaptive migration and optimization for domain characteristics can the classification bottleneck in dedicated scenarios be effectively broken.
However, existing methods still exhibit obvious deficiencies in addressing the service classification problem in urban rail communication scenarios. On one hand, a single deep learning model can hardly simultaneously capture both the structural and statistical features of raw traffic, leading to limited feature expression capability. On the other hand, due to the strong specificity and small sample characteristics of railway services, general network traffic models lack sufficient adaptability in this scenario, making it difficult to balance recognition accuracy, real-time performance, and engineering deployability. Therefore, it is urgent to design an efficient service classification method tailored to the characteristics of railway scenarios. To address the above issues, 
our main contributions can be briefly summarized as below:
\begin{itemize}
    \item We introduce mature deep learning paradigms from general network traffic classification into railway service classification by adopting an image-based traffic representation combined with transfer learning, which learns byte-level texture patterns in a protocol-agnostic manner (bypassing protocol-specific parsing) and mitigates the scarcity of labelled urban-rail samples through knowledge transferred from public general-traffic datasets.
    \item We design a dual-channel cross-modal feature fusion architecture. By jointly learning byte-level textures and flow-level statistical patterns with adaptive channel-attention fusion, it overcomes the representation deficiencies of single-modality models for edge-oriented traffic classification and substantially boosts overall classification performance; as shown in the module-ablation study, it exploits the complementarity of the two branches, which exhibit recognition advantages on different services.
    \item We propose PSL that, while addressing conventional class imbalance, additionally incorporates business-priority weights to enforce stronger gradient penalties on safety-critical services, thereby achieving near-zero-tolerance recall for high-priority classes without sacrificing overall accuracy, ensuring accurate admission of high-priority critical services to downstream urban-rail communication.
\end{itemize}

\section{Related Work}

Focusing on this issue, existing research has undergone an evolutionary process from rule-driven to data-driven approaches. Early methods primarily relied on port-based or protocol-based matching mechanisms~\cite{LIN20091023} to classify services by querying standard ports or analyzing protocol features. However, in scenarios involving dynamic ports and the prevalence of proprietary protocols, their recognition accuracy decreases significantly. Subsequently, DPI methods enhanced recognition capabilities through sophisticated analysis of data payloads~\cite{BUJLOW201575}. Nevertheless, these methods rely on expert experience to construct feature rules, which incurs high computational overhead, and their application is restricted under the constraints of encrypted traffic and privacy protection.

To remedy these drawbacks, machine learning techniques have been widely introduced. In the field of network traffic classification, to enhance model automation and generalization capability, research has gradually shifted towards machine learning-based methods. By extracting statistical features such as packet length distribution and transmission interval, combined with models like Support Vector Machine (SVM)~\cite{abdullah2021machine}, Random Forest~\cite{8756496}, and XGBoost~\cite{article}, service classification is achieved. Although these methods have yielded certain results in structured feature scenarios, their performance is highly dependent on manual feature engineering. Faced with the complexities of diverse service types and significant dynamic traffic changes in railway scenarios, they struggle to achieve stable generalization.

Further progress has been made with the rapid development of deep learning. In recent years, deep learning methods have emerged as a research hotspot in the field of service classification due to their end-to-end feature learning capability. Convolutional Neural Network (CNN)-based models can automatically extract spatial or temporal features from traffic data~\cite{8004872}. Some studies map network traffic to grayscale images and utilize 2D convolutional structures to explore deep patterns. Meanwhile, models such as LSTM and CNN-LSTM~\cite{ULLAH2024190} improve recognition performance by fusing temporal information. Additionally, lightweight networks, autoencoders, and attention mechanisms~\cite{Zheng2022} have been introduced to optimize model efficiency and enhance adaptability to encrypted traffic. Overall, existing research is advancing toward the direction of multi-feature fusion and model intelligence.

More recently, self-supervised pre-training has emerged as a promising paradigm for traffic classification. ET-BERT~\cite{lin2022etbert}pre-trains a Transformer encoder on large-scale unlabeled encrypted traffic using burst-level masked language modeling, learning contextualized datagram representations that generalize across multiple downstream tasks. TrafficFormer~\cite{zhou2025trafficformer}introduces fine-grained pre-training tasks that capture packet direction, order, and flow membership, combined with a data augmentation strategy to reduce reliance on non-semantic protocol fields. YaTC~\cite{zhao2023yetc}recasts traffic classification as a masked image modeling problem, applying a vision Transformer to flow-level feature maps. These pre-training-based methods achieve strong performance on public benchmarks, but their model scales (typically millions of parameters requiring GPU acceleration) are difficult to reconcile with the resource constraints of vehicle-edge hardware. Recent work has shown that carefully designed lightweight convolutional architectures can achieve competitive accuracy with substantially lower computational cost in encrypted traffic classification~\cite{lin2025netconv}, motivating our exploration of a compact CNN-based design for the edge deployment scenario.

Within the railway communication domain, related work on
traffic analysis has primarily focused on intrusion detection.
MSRNet-GLAM~\cite{chen2025msrnet} and Gao et al.~\cite
{gao2021intrusion} propose deep learning and machine learning
methods for detecting anomalous traffic in train communication
networks. These efforts address a binary detection problem,
which differs fundamentally from the multi-class operational
service classification studied here. The present work focuses
on distinguishing normal urban rail service types at the application layer,
an area that remains underexplored.

\begin{figure*}[!tbp]
    \centering
    \includegraphics[trim=0 8 0 8mm, clip, width=\textwidth]{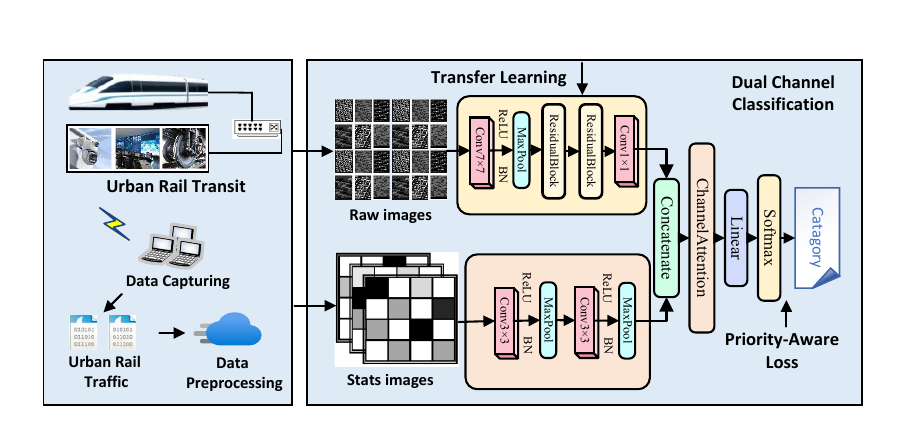}
    \caption{Dual Urban Rail Traffic Classification Method}
    \label{FIG:2}
\end{figure*}

\section{Methodology}

The overall architecture of the proposed model is illustrated in Fig.~\ref{FIG:2}. The raw urban rail traffic is first processed into two distinct image-based representations: the raw traffic channel directly maps the processed byte sequences into a grayscale image to preserve low-level protocol structures, while the statistical feature channel first extracts features such as packet length distribution and inter-arrival times, then concatenates them into a structured statistical block image. These two representations are independently fed into their respective feature extractors. The raw channel employs a modified ResNet initialized with source-domain pre-trained weights to mine deep texture patterns, whereas the statistical channel utilizes a lightweight CNN trained from scratch to capture global statistical distributions. The extracted feature maps are concatenated along the channel dimension and dynamically recalibrated by a channel attention module, which adaptively amplifies the most discriminative representations for each input sample. Finally, the attention-weighted features are passed to a classification head integrated with Dropout, label smoothing, and a priority-sensitive loss function to complete the end-to-end classification process.

\subsection{Raw Channel Feature Extractor}

The raw traffic channel is designed to extract deep texture patterns and protocol structural features from low-level byte sequences. Given the limited sample size of urban rail traffic, this channel employs a modified ResNet architecture integrated with a transfer learning strategy.
Let the input feature map be \(X \in \mathbb{R}^{N \times 1 \times 28 \times 28}\), where \(N\) is the batch size. The feature extraction process of the raw channel is formalized as follows. First, the input passes through a \(7 \times 7\) convolutional layer with stride \(2\) for initial feature extraction and downsampling:

\[
X_1 = \mathrm{ReLU}(\mathrm{BN}(W_{\mathrm{conv1}} * X)) . \tag{1}
\]

At this stage, the spatial dimension is halved, resulting in an output shape of \([N, 64, 14, 14]\). Subsequently, a \(3 \times 3\) max-pooling layer further compresses the spatial information, producing \(X_2 \in \mathbb{R}^{N \times 64 \times 7 \times 7}\). Next, a residual learning stage (Layer 1) consisting of two BasicBlocks is applied. Each residual block introduces a skip connection that adds the input to the convolutional output, mitigating the vanishing gradient problem and extracting high-level semantic features. The output of a residual block can be expressed as:

\[
X_{\mathrm{out}} = F(X_{\mathrm{in}}, W) + X_{\mathrm{in}} . \tag{2}
\]

After Layer 1, the feature map dimension remains \([N, 64, 7, 7]\). Finally, to align with the output of the statistical channel, an adaptive average pooling layer ensures a fixed spatial size, followed by a \(1 \times 1\) convolutional layer (Adjust Conv) for channel-wise adjustment. The final output of the raw channel is \(F_{\mathrm{raw}} \in \mathbb{R}^{N \times 64 \times 7 \times 7}\).

To alleviate the training bottleneck caused by limited labeled samples in the target urban rail domain, a transfer learning strategy is incorporated into the proposed framework, which has been widely adopted for data-scarce scenarios. Considering that different feature modalities exhibit distinct transferability~\cite{5288526}, an asymmetric transfer mechanism is employed. Specifically, transfer learning is applied to the raw traffic channel to facilitate the extraction of low-level texture features. A layer-wise freezing strategy is adopted, where shallow layers (conv1 and bn1) are frozen to retain generic feature representations, while deeper layers remain trainable to adapt high-level semantics to the target domain. In contrast, the statistical channel is trained from scratch to mitigate potential negative transfer. This asymmetric design reflects the distinct transferability of the two modalities. Raw byte images capture transport-layer structural patterns (TCP/IP headers, port numbers, flag fields) that are inherent to the protocol stack and thus largely domain-invariant , which is a property consistent with the transferability of lower-layer CNN features~\cite{yosinski2014transferable}. Statistical features, however, encode application-layer behavioral signatures such as IAT distributions and burstiness profiles, which differ fundamentally between the ISCX source domain (e.g., P2P burst patterns) and the urban rail target domain (e.g., CBTC fixed-period signaling, ERD narrowband voice). Had transfer learning also been applied to the statistical branch, the model would inherit behavioral priors learned from ISCX flows that do not exist in urban rail traffic. These mismatched priors could distort the statistical channel's early representations and degrade, rather than improve, the identification of safety-critical services. Training the statistical branch from scratch therefore preserves the domain-specific patterns essential for distinguishing control services.

\subsection{Statistical Channel Feature Extractor}

Unlike the raw channel which focuses on low-level byte textures, the statistical channel captures global statistical patterns and temporal distribution modes.
The input to this channel is \(X_{\mathrm{stat}} \in \mathbb{R}^{N \times 1 \times 28 \times 28}\). A lightweight CNN with two convolutional blocks extracts spatial features. Each block consists of a \(3 \times 3\) convolution, batch normalization, ReLU, and \(2 \times 2\) max pooling. After two downsampling stages, the network produces the final statistical feature map \(F_{\mathrm{stat}} \in \mathbb{R}^{N \times 64 \times 7 \times 7}\), which aligns with the raw channel output dimension for subsequent fusion.

\subsection{Dual Channel Features Concatenate}

The dual-channel fusion module integrates raw traffic textures and statistical distributions for cross-modal complementarity. We concatenate raw features $F_{\text{raw}} \in \mathbb{R}^{N \times 64 \times 7 \times 7}$ and statistical features $F_{\text{stat}} \in \mathbb{R}^{N \times 64 \times 7 \times 7}$ into $F_{\text{concat}} \in \mathbb{R}^{N \times 128 \times 7 \times 7}$ along the channel dimension.

A channel attention mechanism~\cite{hu2018squeeze} dynamically recalibrates feature weights to enhance discriminative signals and suppress noise.
Finally, the refined features are processed by a classifier with global pooling, Dropout, and a fully connected layer.

\subsection{Loss Function}
\setcounter{equation}{2}  

In general network traffic classification, most existing methods primarily focus on improving classification accuracy and mitigating class imbalance, without explicitly modeling the business priority disparities and QoS heterogeneity inherent in industrial communication scenarios. Cross-entropy loss has been widely adopted in network traffic classification due to its strong stability and effectiveness in probability distribution modeling and classification boundary optimization \cite{RenSE}. In contrast, MSE and $L_2$ based losses are mainly designed for regression tasks and usually exhibit weaker sensitivity to probabilistic outputs, which may lead to slow convergence and insufficient optimization of classification boundaries in multi-class classification scenarios. Although the widely used Focal Loss can alleviate class imbalance by emphasizing hard samples~\cite{lin2017focal}, in urban rail communication systems, different services exhibit significantly different operational criticality and QoS requirements. Combined with the pronounced long-tailed distribution of railway traffic, the standard cross-entropy loss faces a dual challenge in urban rail scenarios: it is prone to gradient dominance by majority classes while failing to impose sufficient misclassification penalties on safety-critical minority services. Therefore, instead of introducing more complex loss structures, this work adopts cross-entropy loss as the optimization foundation and further incorporates two independent weighting coefficients to simultaneously characterize class imbalance and business priority disparities~\cite{elkan2001foundations}. Specifically, the proposed PSL integrates a sample-balancing weight and a business-priority weight, enabling the optimization objective to better align with the practical reliability requirements of urban rail communication systems.

The first coefficient $\omega_c^{\text{prior}}$ explicitly encodes business operational importance and is inversely proportional to the QoS priority value $P_c$:
\begin{equation}
\omega_c^{\text{prior}} = \frac{1/P_c}{\sum_{k=1}^{C} (1/P_k)}.
\end{equation}

The second coefficient $\omega_c^{\text{bal}}$  employs a smoothed inverse-frequency strategy to compensate for sample scarcity and prevent gradient explosion from extreme minority classes:
\begin{equation}
\omega_c^{\text{bal}} = \sqrt{\frac{N_{\text{total}}}{N_c}},
\end{equation}
where $C=5$ is the number of categories, $N_{\text{total}}$ denotes the total training sample count, and $N_c$ is the sample count for class $c$. The final loss function multiplies these two independent factors to jointly optimize training stability and critical-service reliability:
\begin{equation}
\mathcal{L}_{\text{PSL}} = -\frac{1}{N} \sum_{i=1}^{N} \sum_{c=1}^{C} \left( \omega_c^{\text{prior}} \cdot \omega_c^{\text{bal}} \right) y_{i,c} \log(\hat{y}_{i,c}),
\end{equation}
where $N$ is the batch size, $y_{i,c} \in \{0,1\}$ represents the one-hot ground truth label, and $\hat{y}_{i,c}$ is the predicted probability. The coefficient $\omega_c^{\text{prior}}$ ensures that high-priority services receive larger gradient penalties during backpropagation, while $\omega_c^{\text{bal}}$ guarantees sufficient gradient updates for minority classes under long-tailed distributions. Combined with label smoothing, this dual-coefficient optimization objective effectively mitigates overfitting and significantly enhances the model's robustness in classifying critical services under rail environments.Following the service order of Table~\ref{tab:qos_mapping}, the products of the two weights are 0.717, 1.538, 0.231, 0.355, and 0.230. The maximum value (1.538) belongs to ERD, the highest‑priority yet least‑sampled service. The max‑to‑min weight ratio is ~6.7‑fold, lower than the raw sample imbalance of ~11‑fold. These moderate weight magnitudes mitigate potential gradient‑explosion concerns, while jointly realizing sample balance and priority awareness.

\section{Experiments}

In this section, we conduct a series of experiments to comprehensively evaluate the proposed method.
We first describe the datasets and data preprocessing steps (section~\ref{Dataset}), followed by the experiment settings (section~\ref{Experiment Settings}) and evaluation metrics (section~\ref{Evaluation Metrics}).
Then, comparative experiments (section~\ref{Comparison Experiments}) benchmark our method against existing approaches, including rule-based, traditional machine learning, and deep learning models, on a real-world urban rail dataset. These experiments compare not only classification accuracy but also computational complexity and end-to-end latency, demonstrating the engineering feasibility of our method.
Next, we perform two ablation studies: the module ablation study (section~\ref{subsec:ablation_modules}) verifies the individual contributions of the dual-channel architecture, transfer learning, and channel attention mechanism, the loss function ablation study (section~\ref{subsec:loss_ablation}) examines the synergistic effect of the business priority weight and class balance weight in the proposed PSL.

\subsection{Datasets}
\label{Dataset}

In this section, we introduce the two datasets used in our experiments: a public network traffic dataset for transfer learning pre-training and a real-world urban rail dataset for evaluating the proposed method. The following subsections describe the composition, collection process, and key characteristics of each dataset.

{ 
\setlength{\intextsep}{4pt} 
\begin{table}[!tp]
\centering
\footnotesize
\setlength{\abovecaptionskip}{5pt}
\setlength{\belowcaptionskip}{1pt}
\renewcommand{\arraystretch}{1} 
\caption{Composition of the ISCX source-domain dataset with five non-VPN service types}
\label{tab:source_dataset}
\begin{tabularx}{1\linewidth}{c >{\centering\arraybackslash}X}
\toprule
\textbf{Service Type} & \textbf{Representative Applications} \\
\midrule
Chat & Facebook, AIM, ICQ, Skype \\
Email & Outlook, Thunderbird \\
P2P & BitTorrent, Vuze \\
Stream & Netflix, YouTube \\
File & FTP, SMB, SCP \\
\bottomrule
\end{tabularx}
\end{table}
} 

\subsubsection{Source Domain Dataset}
This study utilizes the ISCX VPN-nonVPN dataset as the source domain dataset~\cite{DraperGil2016VPN}, constructed by the Canadian Institute for Cybersecurity (CIC) at the University of New Brunswick in 2016. The complete ISCX VPN-nonVPN collection comprises 14 traffic categories (7 VPN-encrypted and 7 non-VPN). From these, we retain five non-VPN service types (Chat, Email, P2P, Streaming, and File) as the source domain for pre-training. We select only the non-VPN subset because its traffic carries transport-level traffic patterns providing diverse byte-level textures for pre-training the raw channel with the five urban rail services without being altered by an additional VPN encapsulation layer, VPN encryption introduces extra headers and padding fields absent in operational rail traffic. The seven VPN categories are excluded for this reason. These five non-VPN types collectively span a diverse range of communication patterns---interactive messaging, bulk transfer, multimedia streaming, and file exchange---providing a sufficiently broad pre-training basis for the raw channel's transfer learning while remaining visually discriminable in the image domain. The detailed composition is summarized in Table~\ref{tab:source_dataset}.

\subsubsection{Urban Rail Dataset}
To comprehensively validate the effectiveness of the proposed dual-channel model in urban rail vehicle-to-ground communication scenarios, raw PCAP traffic data were collected from a real-world urban rail operational environment over a continuous period.
A dedicated dataset was constructed encompassing five core business categories: CBTC, ERD, Automatic Train Supervision (ATS), Passenger Information Service (PIS), and Closed-Circuit Television (CCTV).
Based on the priority and delay requirements defined in~\cite{10715707}~\cite{zhang2026multi} and the importance of each service as well as the consequences of misclassification, the numerical priority values for the services are determined. The per-class sample counts, functional descriptions, Guaranteed Bit Rate(GBR) resource type and QoS priority mapping are summarized in Table~\ref{tab:qos_mapping}. The dataset is partitioned using a flow-level temporal split strategy: within each service class,all captured flows are sorted chronologically by capture timestamp and divided at a 6:2:2 ratio, with earlier flows used for training and later flows for validation and testing. (This implement replaces the random flow-level split adopted in an earlier version of this study, which could result in temporally correlated traffic flows being distributed across the training and test sets and thus lead to an overly optimistic estimate of generalization performance.) The temporal split provides a stricter evaluation setting by requiring the model to generalize from earlier traffic observations to later unseen traffic patterns, thereby reducing the risk of temporal leakage. All comparative and ablation experiments reported in this study follow this unified temporal splitting protocol.

\begin{figure*}[!tbp]
    \centering
    \vspace{-3mm} 
    \includegraphics[width=0.7\linewidth]{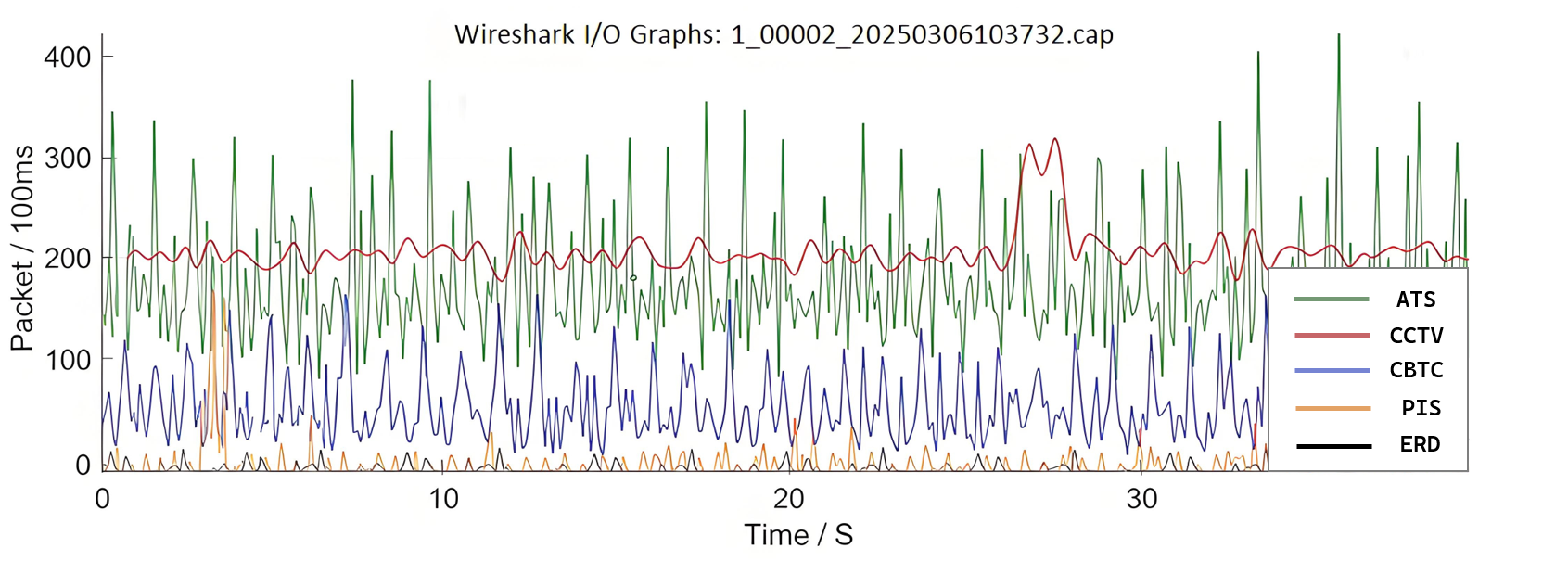}
    \vspace{-2mm} 
    \caption{Real-time I/O throughput curves of five urban rail services by Wireshark}
    \label{fig:wireshark_traffic_analysis}
    \vspace{-4mm} 
\end{figure*}
\begin{table*}[tb]
\centering
\footnotesize
\caption{Urban rail service types, QoS priorities, resource types, and per-class sample counts}
\label{tab:qos_mapping}
\begin{tabular}{ccccc}
\toprule
\textbf{Business Type} & \textbf{Function Description}  & \textbf{Resource Type} & \textbf{Priority ($P_c$)} & \textbf{Sample} \\
\midrule
CBTC & Train Operation Control & GBR & 2 & 1570 \\
ERD & Emergency Radio Dispatch  & GBR & 2 & 341 \\
ATS & Train Status Monitoring  & GBR & 4 & 3776 \\
PIS & Passenger Information Service  & Non-GBR & 6 & 712 \\
CCTV & Platform Video Surveillance  & Non-GBR & 6 & 1692 \\
\bottomrule
\end{tabular}
\end{table*}

Fig.~\ref{fig:wireshark_traffic_analysis} illustrates the real-time I/O throughput curves of these five services captured via Wireshark. Significant modal discrepancies are immediately observable in the temporal domain. As illustrated in the I/O graph, the temporal packet dynamics vary significantly across different services. CCTV (red) maintains a high and stable baseline, reflecting continuous video streaming. CBTC (blue) displays moderate, regular fluctuations, corresponding to periodic control signaling. ATS (green) shows the most intense oscillations, driven by frequent device polling. Meanwhile, PIS (orange) and ERD (black) remain at low baselines with only occasional spikes, matching event-driven transmission. This inherent traffic heterogeneity underscores the limitations of single-feature extraction methods and provides a direct empirical basis for employing the proposed dual-channel fusion architecture in our experimental evaluation.

\subsubsection{Data Preprocessing and Image Mapping}
This section details the preprocessing procedures that convert raw traffic data into image-based representations suitable for deep learning models. Specifically, we present two mapping pipelines: one for raw byte sequences and another for statistical features.

\begin{figure}[!tbp]
\centering
\begin{minipage}[b]{0.15\linewidth}
    \centering
    \vspace{0.3cm}
    \includegraphics[width=\linewidth, height=\linewidth, keepaspectratio]{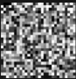}
    \small Chat
\end{minipage}
\hspace{0.1cm}
\begin{minipage}[b]{0.15\linewidth}
    \centering
    \includegraphics[width=\linewidth, height=\linewidth, keepaspectratio]{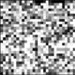}
    \small Email
\end{minipage}
\hspace{0.1cm}
\begin{minipage}[b]{0.15\linewidth}
    \centering
    \includegraphics[width=\linewidth, height=\linewidth, keepaspectratio]{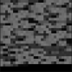}
    \small File
\end{minipage}
\hspace{0.1cm}
\begin{minipage}[b]{0.15\linewidth}
    \centering
    \includegraphics[width=\linewidth, height=\linewidth, keepaspectratio]{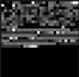}
    \small P2P
\end{minipage}
\hspace{0.1cm}
\begin{minipage}[b]{0.15\linewidth}
    \centering
    \includegraphics[width=\linewidth, height=\linewidth, keepaspectratio]{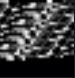}
    \small Stream
\end{minipage}
\caption{Network Traffic Raw Byte Visualization}
\label{fig:traffic_samples1}
\vspace{-3mm}
\end{figure}

\begin{figure}[!tbp]
\centering
\begin{minipage}[b]{0.15\linewidth}
    \centering
    \vspace{0.2cm}
    \includegraphics[width=\linewidth, height=\linewidth, keepaspectratio]{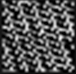}
    \small PIS
\end{minipage}
\hspace{0.1cm}
\begin{minipage}[b]{0.15\linewidth}
    \centering
    \includegraphics[width=\linewidth, height=\linewidth, keepaspectratio]{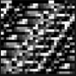}
    \small CCTV
\end{minipage}
\hspace{0.1cm}
\begin{minipage}[b]{0.15\linewidth}
    \centering
    \includegraphics[width=\linewidth, height=\linewidth, keepaspectratio]{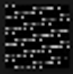}
    \small CBTC
\end{minipage}
\hspace{0.1cm}
\begin{minipage}[b]{0.15\linewidth}
    \centering
    \includegraphics[width=\linewidth, height=\linewidth, keepaspectratio]{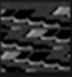}
    \small ERD
\end{minipage}
\hspace{0.1cm}
\begin{minipage}[b]{0.15\linewidth}
    \centering
    \includegraphics[width=\linewidth, height=\linewidth, keepaspectratio]{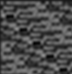}
    \small ATS
\end{minipage}
\caption{Urban Rail Traffic Raw Byte Visualization}
\label{fig:traffic_samples2}
\vspace{-3mm}
\end{figure}

\textit{1) Raw channel:}
To transform one-dimensional traffic data into a two-dimensional format suitable for CNNs, we utilize the USTC-TK-2016 toolkit~\cite{wang2017malware} for standardized processing. First, raw pcap files are split into flows based on 5-tuples and cleaned to remove invalid or abnormal flows. For each processed flow, we truncate or pad the byte sequence to a fixed length and map it into a grayscale image $I \in \mathbb{R}^{1 \times H \times W}$. In this work, the input resolution is fixed at $28 \times 28$. This mapping preserves the spatial local correlation of traffic data, allowing different business categories to exhibit distinct texture distributions in the image domain. As illustrated in Fig.~\ref{fig:traffic_samples1} and Fig.~\ref{fig:traffic_samples2}, both source-domain conventional traffic and urban rail services exhibit distinctly different texture distributions.

\textit{2) Stats Channel:} To comprehensively characterize the temporal, intensity, and distribution properties of urban rail traffic, we extract 16 core statistical features. To adapt these one-dimensional statistical vectors to the input paradigm of deep learning models, we transform them into two-dimensional grayscale images. The specific processing pipeline is as follows: First, specific normalization rules are applied to each statistical feature to linearly compress the raw values into the range \([0, 255]\), eliminating dimensional discrepancies while preserving the relative magnitude of feature distributions. Subsequently, each of the 16 normalized features is mapped into a \(7 \times 7\) feature patch. 

\vspace{-6mm}
\begin{figure}[H]
\centering
\begin{minipage}[b]{0.15\linewidth}
    \centering
    \vspace{0.3cm}
    \includegraphics[width=\linewidth, height=\linewidth, keepaspectratio]{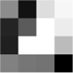}
    \small PIS
\end{minipage}
\hspace{0.1cm}
\begin{minipage}[b]{0.15\linewidth}
    \centering
    \includegraphics[width=\linewidth, height=\linewidth, keepaspectratio]{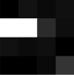}
    \small CCTV
\end{minipage}
\hspace{0.1cm}
\begin{minipage}[b]{0.15\linewidth}
    \centering
    \includegraphics[width=\linewidth, height=\linewidth, keepaspectratio]{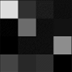}
    \small CBTC
\end{minipage}
\hspace{0.1cm}
\begin{minipage}[b]{0.15\linewidth}
    \centering
    \includegraphics[width=\linewidth, height=\linewidth, keepaspectratio]{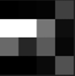}
    \small ERD
\end{minipage}
\hspace{0.1cm}
\begin{minipage}[b]{0.15\linewidth}
    \centering
    \includegraphics[width=\linewidth, height=\linewidth, keepaspectratio]{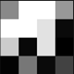}
    \small ATS
\end{minipage}
\caption{Statistical feature image grid (4×4 patches, 16 features) constructed from Urban Rail Traffic}
\label{fig:traffic_samples3}
\vspace{-3mm}
\end{figure}
\vspace{-3mm}

The 16 patches are arranged in a $4 \times 4$ grid according to a manually engineered grouping: Row~1 collects flow-scale statistics (flow duration, total packets, total bytes, average packet size); Row~2 groups rate and boundary statistics (byte rate, packet rate, max packet size, min packet size); Row~3 groups temporal statistics (packet size std, average IAT, IAT std, max IAT); and Row~4 collects shape statistics (burstiness, packet size Q1, packet size Q3, coefficient of variation). The features within each row are statistically related. For example, average IAT and IAT std jointly characterize traffic timing regularity, and max IAT and burstiness capture traffic burst characteristics. These relationships are properties of the features themselves; placing statistically related features in adjacent positions is, however, a manual engineering choice, and the resulting 2-D layout does not reflect an inherent spatial locality of the raw traffic data. This grouping was adopted to align the input dimension with the raw byte channel and to enable a shared downstream pipeline. As illustrated in Fig.~\ref{fig:traffic_samples3}, this transformation intuitively preserves the numerical differences and global distribution patterns of each statistical feature, realizing a cross-modal mapping from one-dimensional statistical information to two-dimensional spatial distribution features.Note that this grouping is tailored to our feature set, and its effectiveness on other datasets is not guaranteed due to potential generalization risk.

\subsection{Experiment Settings}
\label{Experiment Settings}

The training hyperparameters are configured as follows. The batch size is set to 16. To accommodate modality-specific convergence dynamics, the initial learning rate for the transfer learning branch in the raw channel is set to $5 \times 10^{-5}$, while the statistical branch, attention module, and classification head share a unified rate of $2 \times 10^{-4}$. Weight decay is fixed at $6 \times 10^{-4}$, and dropout rates of 0.2 and 0.1 are applied to the fully connected layers of the classification head to mitigate overfitting. Regarding the implementation environment, all models are developed in PyTorch 1.13.1 and accelerated by CUDA 11.6 with cuDNN 8.3.0. All experiments are conducted on a workstation equipped with an Intel Core i7 processor, 8\,GB RAM, and an NVIDIA GeForce MX250 GPU (2\,GB VRAM).

\subsection{Evaluation Metrics}
\label{Evaluation Metrics}
\setcounter{equation}{5}

All deep-learning experiments are repeated over five independent random seeds---(42, 123, 2024, 7, 99)---that jointly control weight initialization, training-data shuffling order, and dropout mask generation. All performance metrics are reported in the format $\text{mean} \pm \text{standard deviation}$ across the five runs. The final confusion matrix presented in each experiment is selected as the seed whose accuracy is closest to the five-seed mean.

For non-iterative methods (Port-rule and DPI), the reported standard deviation reflects inherent classification stability across seeds rather than training variability. The five seeds are fixed identically across all compared methods and ablation configurations.

To evaluate model performance against existing traffic classification methods, we adopt Accuracy, Precision, Recall, and F1-score as baseline metrics \cite{RenAD}. Standard macro-averaging treats all classes equally, which fails to reflect the classification performance disparity of safety-critical services in heterogeneous urban rail traffic. To address this, we replace macro-averaging with a priority-weighted mechanism. Per-class metrics are first defined as:

\begin{equation}
Pre_c = \frac{TP_c}{TP_c + FP_c} ,
\end{equation}
\begin{equation}
Rec_c = \frac{TP_c}{TP_c + FN_c} ,
\end{equation}
\begin{equation}
F1_c = 2 \times \frac{Pre_c \times Rec_c}{Pre_c + Rec_c} ,
\end{equation}
where \(TP_c\), \(FP_c\), \(FN_c\) denote the true positives, false positives, and false negatives for class \(c\), respectively. These are then aggregated using the normalized priority weight \(\omega_c^{\mathrm{prior}}\) (inversely proportional to QoS priority \(P_c\)):

\begin{equation}
W\text{-}X = \sum_{c=1}^{C} \omega_c^{\mathrm{prior}} \cdot X_c, \quad X \in \{Pre, Rec, F1\} .
\end{equation}

Overall accuracy is computed as the ratio of correctly classified samples to total samples. Crucially, unlike the loss function which incorporates a sample-balance term \(\omega_c^{\mathrm{bal}}\) for training stability, the evaluation metrics strictly utilize \(\omega_c^{\mathrm{prior}}\) alone. This separation is principled. During training, the optimizer must simultaneously combat the severe class imbalance and prioritize safety-critical services; the product form ensures neither objective is subordinated. During evaluation, the metric must answer that how well are operational safety requirements satisfied? Introducing $\omega_{bal}^c$ into evaluation would inflate the importance of minority classes beyond their actual operational criticality. By restricting evaluation weights to $\omega_{prior}^c$ alone, the reported metrics directly quantify QoS compliance. To ensure fair comparison with methods not explicitly optimized for weighted objectives in comparison experiments, we additionally report standard Macro data, which treats all five classes with equal weight, alongside the priority-weighted metrics in all result tables.

\subsection{Comparison Experiments}
\label{Comparison Experiments}

To comprehensively evaluate how well-established network traffic classification methods perform when applied to the urban rail service traffic domain, we conduct a series of comparison experiments on the real-world urban rail dataset, including both traditional methods and advanced deep learning models. This section describes the experimental design, including baseline methods and input paradigms, followed by detailed analysis of classification performance and real-time feasibility.

\subsubsection{Comparison Experiments Design}
Traditional methods include: (1) port-based rule matching (Port-rule), (2) DPI, and (3) Random Forest based on statistical features~\cite{ELMAGHRABY2024102361}. These baselines highlight the inherent limitations of rule-driven and simple statistical models when facing real urban-rail transit environment.

For deep learning baselines, we select four representative methods and strictly reproduce them according to the input formats and preprocessing pipelines described in their original papers: (1) One-Dimensional Convolutional Neural Network (1D-CNN)~\cite{8004872}, which takes the first 784 raw bytes of each session as input, requiring flow splitting, cleaning, and fixed-length truncation or zero-padding; (2) CNN-LSTM~\cite{ULLAH2024190}, which takes 49 statistical features (e.g., mean packet length, variance, inter-arrival time) extracted from network flows as input, requiring per-flow statistical calculation and normalization; (3) Graph Neural Network (GNN)~\cite{9979671}, which takes graph-structured data as input, where each node represents 1,500 raw bytes of a packet, edges represent temporal relationships, and global attributes include five flow-level statistical features, requiring each flow to be mapped to a directed graph; (4) Multi-Task Transformer (MTT)~\cite{Zheng2022}, which takes 1,500 raw bytes of a single packet (Ethernet header removed, IP addresses zeroed, TCP/UDP headers unified to 20 bytes) as input, fed directly as a byte sequence without flow reassembly.

The input paradigms directly affect end-to-end latency. Methods such as 1D-CNN and MTT directly ingest raw byte sequences, bypassing feature engineering and format conversion steps, which gives them a latency advantage. In contrast, CNN-LSTM, GNN, and our method incorporate additional data processing pipelines (e.g., sequence framing, graph topology construction, or statistical feature mapping and image generation), which inevitably introduce preprocessing overhead but provide more discriminative features.

All compared models are trained and tested under the same dataset and hardware environment to ensure fairness. Detailed performance comparisons are presented in Table~\ref{tab:comparison}, while the computational complexity and end-to-end latency of each model are summarized in Fig.~\ref{figa:single_fig} and Fig.~\ref{figc:single_fig}.

\begin{table*}[tb]
\centering
\scriptsize
\caption{Performance comparison across traditional, ML, and DL methods on the urban rail dataset (five-seed mean±std)}
\label{tab:comparison}
\setlength{\tabcolsep}{2.6pt}
\renewcommand{\arraystretch}{1.18}
\newcommand{\cellpad}{\strut}
\begin{tabularx}{\textwidth}{
>{\centering\arraybackslash}c
|>{\centering\arraybackslash}c
|>{\centering\arraybackslash}c
|>{\centering\arraybackslash}c
|>{\centering\arraybackslash}X
|>{\centering\arraybackslash}X
|>{\centering\arraybackslash}X
|>{\centering\arraybackslash}X
|>{\centering\arraybackslash}X
|>{\centering\arraybackslash}X
|>{\centering\arraybackslash}X
}
\hline
\textbf{Method} & \textbf{Input} & \textbf{Acc} & \textbf{Metric} & \textbf{PIS} & \textbf{CCTV} & \textbf{CBTC} & \textbf{ERD} & \textbf{ATS} & \textbf{Macro} & \textbf{Weighted} \\
\hline\noalign{\vskip 3pt}

\multirow{3}{*}{Port-rule}
& \multirow{3}{*}{Raw Bytes}
& \multirow{3}{*}{\shortstack[c]{\cellpad 83.36 \\ ($\pm$ 0.04)\cellpad}}
& Recall
& \shortstack[c]{\cellpad 76.90 \\ ($\pm$ 0.28)\cellpad}
& \shortstack[c]{\cellpad 77.81 \\ ($\pm$ 0.19)\cellpad}
& \shortstack[c]{\cellpad 83.57 \\ ($\pm$ 0.32)\cellpad}
& \shortstack[c]{\cellpad 73.24 \\ ($\pm$ 0.59)\cellpad}
& \shortstack[c]{\cellpad 87.89 \\ ($\pm$ 0.06)\cellpad}
& \shortstack[c]{\cellpad 79.88 \\ ($\pm$ 0.11)\cellpad}
& \shortstack[c]{\cellpad 79.68 \\ ($\pm$ 0.13)\cellpad} \\
& & & Precision
& \shortstack[c]{\cellpad 59.71 \\ ($\pm$ 1.51)\cellpad}
& \shortstack[c]{\cellpad 79.36 \\ ($\pm$ 1.94)\cellpad}
& \shortstack[c]{\cellpad 92.27 \\ ($\pm$ 0.89)\cellpad}
& \shortstack[c]{\cellpad 59.92 \\ ($\pm$ 2.20)\cellpad}
& \shortstack[c]{\cellpad 90.31 \\ ($\pm$ 0.61)\cellpad}
& \shortstack[c]{\cellpad 76.32 \\ ($\pm$ 0.35)\cellpad}
& \shortstack[c]{\cellpad 76.96 \\ ($\pm$ 0.69)\cellpad} \\
& & & F1-score
& \shortstack[c]{\cellpad 67.21 \\ ($\pm$ 1.03)\cellpad}
& \shortstack[c]{\cellpad 78.57 \\ ($\pm$ 0.85)\cellpad}
& \shortstack[c]{\cellpad 87.70 \\ ($\pm$ 0.38)\cellpad}
& \shortstack[c]{\cellpad 65.90 \\ ($\pm$ 1.54)\cellpad}
& \shortstack[c]{\cellpad 89.09 \\ ($\pm$ 0.29)\cellpad}
& \shortstack[c]{\cellpad 77.69 \\ ($\pm$ 0.28)\cellpad}
& \shortstack[c]{\cellpad 77.92 \\ ($\pm$ 0.46)\cellpad} \\
\hline

\multirow{3}{*}{DPI}
& \multirow{3}{*}{Raw Bytes}
& \multirow{3}{*}{\shortstack[c]{\cellpad 91.06 \\ ($\pm$ 0.05)\cellpad}}
& Recall
& \shortstack[c]{\cellpad 89.86 \\ ($\pm$ 0.34)\cellpad}
& \shortstack[c]{\cellpad 88.99 \\ ($\pm$ 0.12)\cellpad}
& \shortstack[c]{\cellpad 95.16 \\ ($\pm$ 0.34)\cellpad}
& \shortstack[c]{\cellpad 89.12 \\ ($\pm$ 0.72)\cellpad}
& \shortstack[c]{\cellpad 90.68 \\ ($\pm$ 0.06)\cellpad}
& \shortstack[c]{\cellpad 90.76 \\ ($\pm$ 0.09)\cellpad}
& \shortstack[c]{\cellpad 91.34 \\ ($\pm$ 0.15)\cellpad} \\
& & & Precision
& \shortstack[c]{\cellpad 75.53 \\ ($\pm$ 1.55)\cellpad}
& \shortstack[c]{\cellpad 87.62 \\ ($\pm$ 1.49)\cellpad}
& \shortstack[c]{\cellpad 95.78 \\ ($\pm$ 0.87)\cellpad}
& \shortstack[c]{\cellpad 76.03 \\ ($\pm$ 2.62)\cellpad}
& \shortstack[c]{\cellpad 96.05 \\ ($\pm$ 0.45)\cellpad}
& \shortstack[c]{\cellpad 86.20 \\ ($\pm$ 0.42)\cellpad}
& \shortstack[c]{\cellpad 86.59 \\ ($\pm$ 0.73)\cellpad} \\
& & & F1-score
& \shortstack[c]{\cellpad 82.07 \\ ($\pm$ 0.98)\cellpad}
& \shortstack[c]{\cellpad 88.30 \\ ($\pm$ 0.70)\cellpad}
& \shortstack[c]{\cellpad 95.47 \\ ($\pm$ 0.38)\cellpad}
& \shortstack[c]{\cellpad 82.03 \\ ($\pm$ 1.55)\cellpad}
& \shortstack[c]{\cellpad 93.28 \\ ($\pm$ 0.19)\cellpad}
& \shortstack[c]{\cellpad 88.23 \\ ($\pm$ 0.25)\cellpad}
& \shortstack[c]{\cellpad 88.71 \\ ($\pm$ 0.42)\cellpad} \\
\hline

\multirow{3}{*}{\shortstack[c]{Random\\Forest\\\cite{ELMAGHRABY2024102361}}}
& \multirow{3}{*}{Flow Features}
& \multirow{3}{*}{\shortstack[c]{\cellpad 88.81 \\ ($\pm$ 0.14)\cellpad}}
& Recall
& \shortstack[c]{\cellpad 88.31 \\ ($\pm$ 1.05)\cellpad}
& \shortstack[c]{\cellpad 86.92 \\ ($\pm$ 0.43)\cellpad}
& \shortstack[c]{\cellpad 92.29 \\ ($\pm$ 0.74)\cellpad}
& \shortstack[c]{\cellpad 88.24 \\ ($\pm$ 1.61)\cellpad}
& \shortstack[c]{\cellpad 88.34 \\ ($\pm$ 0.19)\cellpad}
& \shortstack[c]{\cellpad 88.82 \\ ($\pm$ 0.38)\cellpad}
& \shortstack[c]{\cellpad 89.40 \\ ($\pm$ 0.39)\cellpad} \\
& & & Precision
& \shortstack[c]{\cellpad 70.87 \\ ($\pm$ 1.48)\cellpad}
& \shortstack[c]{\cellpad 84.80 \\ ($\pm$ 1.56)\cellpad}
& \shortstack[c]{\cellpad 94.97 \\ ($\pm$ 0.92)\cellpad}
& \shortstack[c]{\cellpad 71.00 \\ ($\pm$ 2.82)\cellpad}
& \shortstack[c]{\cellpad 94.80 \\ ($\pm$ 0.61)\cellpad}
& \shortstack[c]{\cellpad 83.29 \\ ($\pm$ 0.50)\cellpad}
& \shortstack[c]{\cellpad 83.77 \\ ($\pm$ 0.83)\cellpad} \\
& & & F1-score
& \shortstack[c]{\cellpad 78.63 \\ ($\pm$ 1.20)\cellpad}
& \shortstack[c]{\cellpad 85.84 \\ ($\pm$ 0.62)\cellpad}
& \shortstack[c]{\cellpad 93.61 \\ ($\pm$ 0.51)\cellpad}
& \shortstack[c]{\cellpad 78.67 \\ ($\pm$ 2.16)\cellpad}
& \shortstack[c]{\cellpad 91.46 \\ ($\pm$ 0.24)\cellpad}
& \shortstack[c]{\cellpad 85.64 \\ ($\pm$ 0.42)\cellpad}
& \shortstack[c]{\cellpad 86.15 \\ ($\pm$ 0.59)\cellpad} \\
\hline

\multirow{3}{*}{1D-CNN~\cite{8004872}}
& \multirow{3}{*}{Raw Bytes}
& \multirow{3}{*}{\shortstack[c]{\cellpad 89.56 \\ ($\pm$ 0.24)\cellpad}}
& Recall
& \shortstack[c]{\cellpad 86.76 \\ ($\pm$ 1.57)\cellpad}
& \shortstack[c]{\cellpad 83.37 \\ ($\pm$ 0.76)\cellpad}
& \shortstack[c]{\cellpad 95.99 \\ ($\pm$ 0.82)\cellpad}
& \shortstack[c]{\cellpad 93.53 \\ ($\pm$ 2.20)\cellpad}
& \shortstack[c]{\cellpad 89.83 \\ ($\pm$ 0.23)\cellpad}
& \shortstack[c]{\cellpad 89.90 \\ ($\pm$ 0.57)\cellpad}
& \shortstack[c]{\cellpad 91.94 \\ ($\pm$ 0.61)\cellpad} \\
& & & Precision
& \shortstack[c]{\cellpad 69.95 \\ ($\pm$ 1.91)\cellpad}
& \shortstack[c]{\cellpad 85.80 \\ ($\pm$ 1.60)\cellpad}
& \shortstack[c]{\cellpad 95.69 \\ ($\pm$ 0.88)\cellpad}
& \shortstack[c]{\cellpad 73.84 \\ ($\pm$ 2.62)\cellpad}
& \shortstack[c]{\cellpad 95.39 \\ ($\pm$ 0.59)\cellpad}
& \shortstack[c]{\cellpad 84.14 \\ ($\pm$ 0.49)\cellpad}
& \shortstack[c]{\cellpad 84.99 \\ ($\pm$ 0.79)\cellpad} \\
& & & F1-score
& \shortstack[c]{\cellpad 77.45 \\ ($\pm$ 1.59)\cellpad}
& \shortstack[c]{\cellpad 84.55 \\ ($\pm$ 0.48)\cellpad}
& \shortstack[c]{\cellpad 95.83 \\ ($\pm$ 0.52)\cellpad}
& \shortstack[c]{\cellpad 82.51 \\ ($\pm$ 2.23)\cellpad}
& \shortstack[c]{\cellpad 92.52 \\ ($\pm$ 0.23)\cellpad}
& \shortstack[c]{\cellpad 86.57 \\ ($\pm$ 0.50)\cellpad}
& \shortstack[c]{\cellpad 87.98 \\ ($\pm$ 0.64)\cellpad} \\
\hline

\multirow{3}{*}{CNN-LSTM~\cite{ULLAH2024190}}
& \multirow{3}{*}{Flow Features}
& \multirow{3}{*}{\shortstack[c]{\cellpad 92.62 \\ ($\pm$ 0.09)\cellpad}}
& Recall
& \shortstack[c]{\cellpad 89.86 \\ ($\pm$ 0.56)\cellpad}
& \shortstack[c]{\cellpad 88.34 \\ ($\pm$ 0.40)\cellpad}
& \shortstack[c]{\cellpad 97.39 \\ ($\pm$ 0.47)\cellpad}
& \shortstack[c]{\cellpad 95.88 \\ ($\pm$ 1.10)\cellpad}
& \shortstack[c]{\cellpad 92.77 \\ ($\pm$ 0.14)\cellpad}
& \shortstack[c]{\cellpad 92.85 \\ ($\pm$ 0.27)\cellpad}
& \shortstack[c]{\cellpad 94.44 \\ ($\pm$ 0.28)\cellpad} \\
& & & Precision
& \shortstack[c]{\cellpad 77.46 \\ ($\pm$ 1.84)\cellpad}
& \shortstack[c]{\cellpad 89.97 \\ ($\pm$ 1.41)\cellpad}
& \shortstack[c]{\cellpad 96.90 \\ ($\pm$ 0.83)\cellpad}
& \shortstack[c]{\cellpad 80.80 \\ ($\pm$ 3.21)\cellpad}
& \shortstack[c]{\cellpad 96.77 \\ ($\pm$ 0.44)\cellpad}
& \shortstack[c]{\cellpad 88.38 \\ ($\pm$ 0.58)\cellpad}
& \shortstack[c]{\cellpad 89.02 \\ ($\pm$ 0.93)\cellpad} \\
& & & F1-score
& \shortstack[c]{\cellpad 83.19 \\ ($\pm$ 1.24)\cellpad}
& \shortstack[c]{\cellpad 89.14 \\ ($\pm$ 0.49)\cellpad}
& \shortstack[c]{\cellpad 97.14 \\ ($\pm$ 0.45)\cellpad}
& \shortstack[c]{\cellpad 87.67 \\ ($\pm$ 2.16)\cellpad}
& \shortstack[c]{\cellpad 94.73 \\ ($\pm$ 0.17)\cellpad}
& \shortstack[c]{\cellpad 90.37 \\ ($\pm$ 0.43)\cellpad}
& \shortstack[c]{\cellpad 91.46 \\ ($\pm$ 0.62)\cellpad} \\
\hline

\multirow{3}{*}{GNN~\cite{9979671}}
& \multirow{3}{*}{Flow Graph}
& \multirow{3}{*}{\shortstack[c]{\cellpad 91.09 \\ ($\pm$ 0.19)\cellpad}}
& Recall
& \shortstack[c]{\cellpad 88.45 \\ ($\pm$ 1.31)\cellpad}
& \shortstack[c]{\cellpad 85.92 \\ ($\pm$ 0.66)\cellpad}
& \shortstack[c]{\cellpad 97.45 \\ ($\pm$ 0.67)\cellpad}
& \shortstack[c]{\cellpad 94.41 \\ ($\pm$ 1.71)\cellpad}
& \shortstack[c]{\cellpad 90.97 \\ ($\pm$ 0.18)\cellpad}
& \shortstack[c]{\cellpad 91.44 \\ ($\pm$ 0.48)\cellpad}
& \shortstack[c]{\cellpad 93.31 \\ ($\pm$ 0.50)\cellpad} \\
& & & Precision
& \shortstack[c]{\cellpad 73.74 \\ ($\pm$ 1.91)\cellpad}
& \shortstack[c]{\cellpad 87.77 \\ ($\pm$ 1.49)\cellpad}
& \shortstack[c]{\cellpad 96.23 \\ ($\pm$ 0.76)\cellpad}
& \shortstack[c]{\cellpad 77.23 \\ ($\pm$ 2.56)\cellpad}
& \shortstack[c]{\cellpad 96.14 \\ ($\pm$ 0.55)\cellpad}
& \shortstack[c]{\cellpad 86.22 \\ ($\pm$ 0.51)\cellpad}
& \shortstack[c]{\cellpad 86.96 \\ ($\pm$ 0.76)\cellpad} \\
& & & F1-score
& \shortstack[c]{\cellpad 80.42 \\ ($\pm$ 1.49)\cellpad}
& \shortstack[c]{\cellpad 86.82 \\ ($\pm$ 0.47)\cellpad}
& \shortstack[c]{\cellpad 96.84 \\ ($\pm$ 0.46)\cellpad}
& \shortstack[c]{\cellpad 84.94 \\ ($\pm$ 1.92)\cellpad}
& \shortstack[c]{\cellpad 93.48 \\ ($\pm$ 0.20)\cellpad}
& \shortstack[c]{\cellpad 88.50 \\ ($\pm$ 0.46)\cellpad}
& \shortstack[c]{\cellpad 89.77 \\ ($\pm$ 0.57)\cellpad} \\
\hline

\multirow{3}{*}{MTT~\cite{Zheng2022}}
& \multirow{3}{*}{Raw Bytes}
& \multirow{3}{*}{\shortstack[c]{\cellpad 95.16 \\ ($\pm$ 0.06)\cellpad}}
& Recall
& \shortstack[c]{\cellpad 94.23 \\ ($\pm$ 0.53)\cellpad}
& \shortstack[c]{\cellpad 91.42 \\ ($\pm$ 0.19)\cellpad}
& \shortstack[c]{\cellpad 98.92 \\ ($\pm$ 0.25)\cellpad}
& \shortstack[c]{\cellpad 98.24 \\ ($\pm$ 0.59)\cellpad}
& \shortstack[c]{\cellpad 95.18 \\ ($\pm$ 0.06)\cellpad}
& \shortstack[c]{\cellpad 95.60 \\ ($\pm$ 0.13)\cellpad}
& \shortstack[c]{\cellpad 96.83 \\ ($\pm$ 0.13)\cellpad} \\
& & & Precision
& \shortstack[c]{\cellpad 83.86 \\ ($\pm$ 1.52)\cellpad}
& \shortstack[c]{\cellpad 93.48 \\ ($\pm$ 1.09)\cellpad}
& \shortstack[c]{\cellpad 98.17 \\ ($\pm$ 0.69)\cellpad}
& \shortstack[c]{\cellpad 86.58 \\ ($\pm$ 2.24)\cellpad}
& \shortstack[c]{\cellpad 98.01 \\ ($\pm$ 0.31)\cellpad}
& \shortstack[c]{\cellpad 92.02 \\ ($\pm$ 0.30)\cellpad}
& \shortstack[c]{\cellpad 92.49 \\ ($\pm$ 0.62)\cellpad} \\
& & & F1-score
& \shortstack[c]{\cellpad 88.73 \\ ($\pm$ 0.96)\cellpad}
& \shortstack[c]{\cellpad 92.43 \\ ($\pm$ 0.44)\cellpad}
& \shortstack[c]{\cellpad 98.54 \\ ($\pm$ 0.33)\cellpad}
& \shortstack[c]{\cellpad 92.03 \\ ($\pm$ 1.33)\cellpad}
& \shortstack[c]{\cellpad 96.57 \\ ($\pm$ 0.14)\cellpad}
& \shortstack[c]{\cellpad 93.66 \\ ($\pm$ 0.19)\cellpad}
& \shortstack[c]{\cellpad 94.50 \\ ($\pm$ 0.35)\cellpad} \\
\hline

\multirow{3}{*}{Ours}
& \multirow{3}{*}{\shortstack[c]{\cellpad Raw Bytes \\ + Flow Features\cellpad}}
& \multirow{3}{*}{\shortstack[c]{\cellpad 98.74 \\ ($\pm$ 0.05)\cellpad}}
& Recall
& \shortstack[c]{\cellpad 98.73 \\ ($\pm$ 0.28)\cellpad}
& \shortstack[c]{\cellpad 98.76 \\ ($\pm$ 0.12)\cellpad}
& \shortstack[c]{\cellpad 99.94 \\ ($\pm$ 0.13)\cellpad}
& \shortstack[c]{\cellpad 99.41 \\ ($\pm$ 0.72)\cellpad}
& \shortstack[c]{\cellpad 98.17 \\ ($\pm$ 0.05)\cellpad}
& \shortstack[c]{\cellpad 99.00 \\ ($\pm$ 0.17)\cellpad}
& \shortstack[c]{\cellpad 99.24 \\ ($\pm$ 0.23)\cellpad} \\
& & & Precision
& \shortstack[c]{\cellpad 95.91 \\ ($\pm$ 0.93)\cellpad}
& \shortstack[c]{\cellpad 97.83 \\ ($\pm$ 0.39)\cellpad}
& \shortstack[c]{\cellpad 99.37 \\ ($\pm$ 0.44)\cellpad}
& \shortstack[c]{\cellpad 96.31 \\ ($\pm$ 1.08)\cellpad}
& \shortstack[c]{\cellpad 99.68 \\ ($\pm$ 0.14)\cellpad}
& \shortstack[c]{\cellpad 97.82 \\ ($\pm$ 0.21)\cellpad}
& \shortstack[c]{\cellpad 97.93 \\ ($\pm$ 0.31)\cellpad} \\
& & & F1-score
& \shortstack[c]{\cellpad 97.30 \\ ($\pm$ 0.50)\cellpad}
& \shortstack[c]{\cellpad 98.29 \\ ($\pm$ 0.15)\cellpad}
& \shortstack[c]{\cellpad 99.65 \\ ($\pm$ 0.23)\cellpad}
& \shortstack[c]{\cellpad 97.83 \\ ($\pm$ 0.64)\cellpad}
& \shortstack[c]{\cellpad 98.92 \\ ($\pm$ 0.06)\cellpad}
& \shortstack[c]{\cellpad 98.40 \\ ($\pm$ 0.14)\cellpad}
& \shortstack[c]{\cellpad 98.57 \\ ($\pm$ 0.21)\cellpad} \\
\hline
\end{tabularx}
\vspace{-10pt}
\end{table*}

\subsubsection{Results Analysis}
Table~\ref{tab:comparison} reports the five-seed mean $\pm$ standard deviation of all methods under the temporal split. The results not only reflect differences in overall accuracy, but more fundamentally reveal the capability gap among various approaches in handling the intrinsic characteristics of railway traffic, including multi-service concurrency, heterogeneous QoS requirements, and asymmetric misclassification costs.

Traditional rule-based approaches (Port-rule and DPI), which involve no stochastic training, the reported standard deviations reflect the intrinsic stability of their deterministic decision rules over the fixed split, rather than any training randomness, exhibit clear limitations in this scenario. Port-rule achieves only $83.36 (\pm0.04)\%$ accuracy and DPI reaches $91.06 (\pm0.05)\%$, both with negligible standard deviation,which is an expected consequence of their reliance on fixed protocol signatures rather than learned representations. Their recall on safety-critical services remains inadequate: DPI identifies $89.12\%$ of ERD flows, and Port-rule drops to $73.24\%$, which is insufficient for reliable detection of safety-critical services. Although Random Forest, trained on handcrafted statistical features, improves accuracy to $88.81 (\pm0.14)\%$, its relatively large standard deviation, the highest among non-deep methods, indicates that manual feature engineering alone cannot provide stable classification under temporal traffic variation.

Deep learning methods alleviate these limitations through end-to-end feature learning, but their performance diverges sharply depending on the input modality. Raw-byte methods (1D-CNN and MTT) achieve strong recall on services with fixed packet structures---MTT reaches $98.92\%$ on CBTC---by learning discriminative byte-level patterns directly from payload sequences. However, their recall on variable-content services degrades significantly: 1D-CNN attains only $83.37\%$ on CCTV, where scene-dependent video encoding produces byte textures that differ markedly across time periods. Conversely, flow-feature methods (CNN-LSTM and GNN) capture statistical regularities at the session level, such as packet-length distributions and inter-arrival-time statistics, but lack access to the structural cues embedded in raw payloads. As a result, their CBTC recall ($97.39\%$ for CNN-LSTM and $97.45\%$ for GNN) consistently trails MTT's, while their ERD recall remains in the range of $94\%$--$96\%$. For non-critical services with larger sample sizes, such as ATS and PIS, all deep baselines perform acceptably, with recall above $88\%$, reflecting the fact that these classes are less sensitive to modality limitations and temporal variation. The accuracy standard deviations reinforce these observations: 1D-CNN ($\pm0.24\%$) and GNN ($\pm0.19\%$) are markedly less stable than MTT ($\pm0.06\%$) and CNN-LSTM ($\pm0.09\%$), confirming that weaker single-modality models are more vulnerable to random initialization and data partitioning. The complementary strengths of the two modalities, raw-byte methods excelling on fixed-structure control services, flow-feature methods better handling statistical regularity, directly motivate the dual-channel design of our proposed framework.

The proposed method achieves $98.74 (\pm0.05)\%$ accuracy, with recall of $99.94\%$ on CBTC and $99.41\%$ on ERD, and maintains precision above $95.9\%$ across all five service categories. The reported performance gain cannot be attributed to the dual-channel architecture alone, but arises from the joint effect of the dual-channel architecture and the PSL, with the two components contributing to a comparable degree (as can be inferred from the ablation studies in Sections~\ref{subsec:ablation_modules} and~\ref{subsec:loss_ablation}: with the loss fixed, the architecture improves accuracy by $4.23$~pp over the best single-channel variant, $94.51\%$ to $98.74\%$; with the architecture fixed, the PSL improves accuracy by $5.13$~pp over cross-entropy training, $93.61\%$ to $98.74\%$). Notably, under matched cross-entropy training the dual-channel architecture alone reaches $93.61\%$, slightly below the strongest baseline MTT ($95.16\%$); neither component alone therefore surpasses the strongest baseline.The raw channel, enhanced by transfer learning from the ISCX VPN source domain, supplies fine-grained byte-level textures that prove effective for multimedia services; the statistical channel captures macroscopic temporal patterns such as inter-arrival times and burstiness profiles, which are critical for distinguishing periodic control services. The channel attention mechanism further adaptively fuses these complementary representations, suppressing redundant information and amplifying discriminative features on a per-sample basis. The $\pm0.05\%$ accuracy standard deviation---the smallest among all trained models---confirms that this multi-modal architecture yields robust convergence, with the two channels providing redundant yet complementary evidence that stabilizes predictions across different random seeds.

Under priority-weighted evaluation, the proposed method achieves a Weighted Recall of $0.9924(\pm0.0023)$ and a Weighted F1 of $0.9857 (\pm0.0021)$, exceeding MTT's $0.9683$ and $0.9450$, respectively. The standard Macro-F1, which treats all five classes with equal weight, reaches $0.9840$, a margin of $0.0474$ over MTT ($0.9366$), confirming that the advantage also holds under the unweighted macro metric and does not depend on the priority weighting; the weighted metrics are therefore presented as a complementary view that highlights the model's behavior on safety-critical services. From an engineering standpoint, this level of accuracy on safety-critical control traffic, achieved with only $0.18$M parameters, enables reliable automated service differentiation at the network edge without manual rule configuration, laying the groundwork for downstream tasks such as per-service QoS scheduling and dynamic resource allocation in urban rail communication systems.

\begin{figure}[tbp]
    \centering
    \includegraphics[width=1\linewidth]{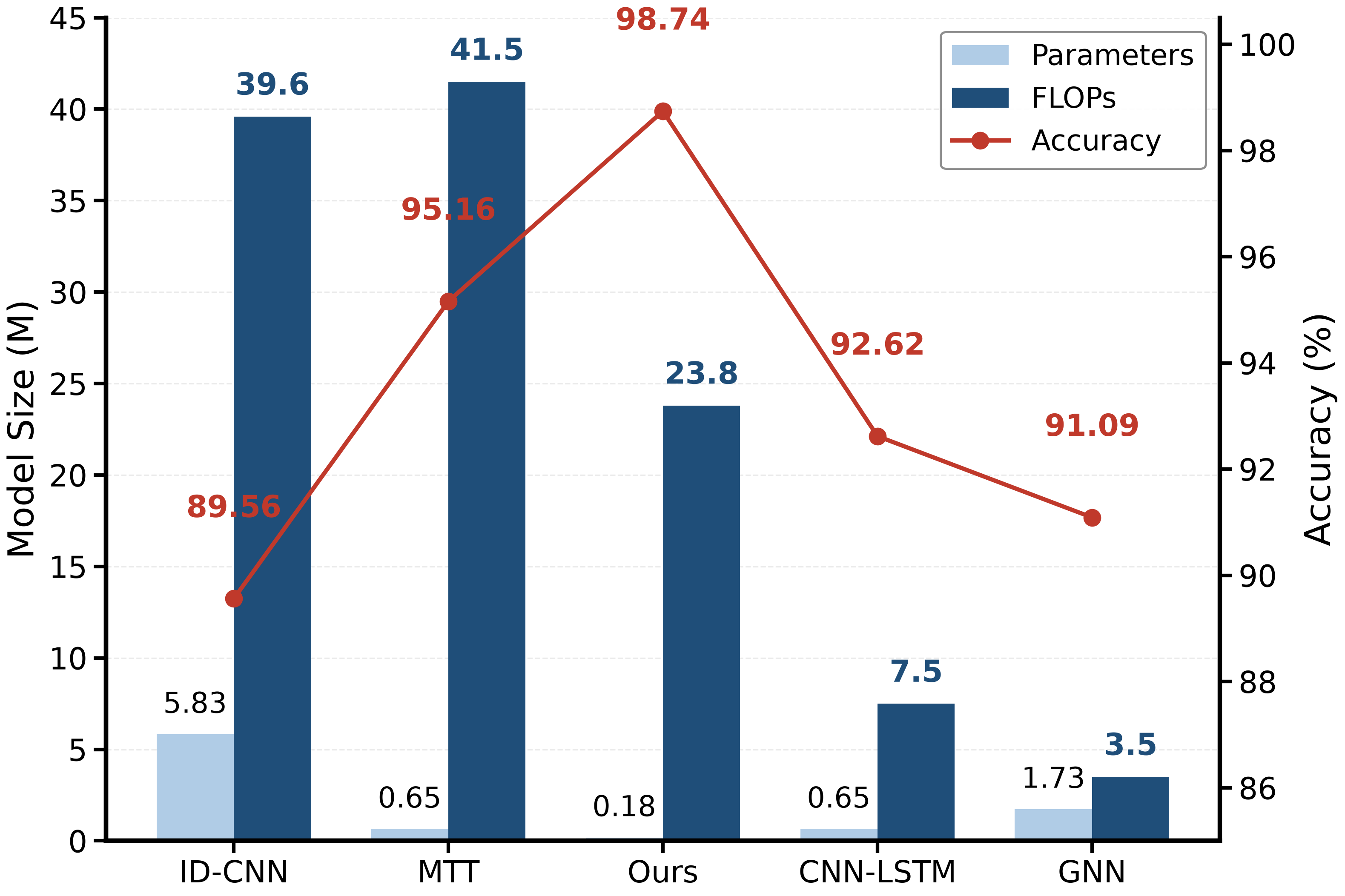} 
    \caption{Accuracy vs model complexity comparison: Parameters , FLOPs , and accuracy across five methods on the urban rail dataset.}
    \label{figa:single_fig}
\end{figure}

\begin{figure}[tbp]
    \centering
    \includegraphics[width=1\linewidth]{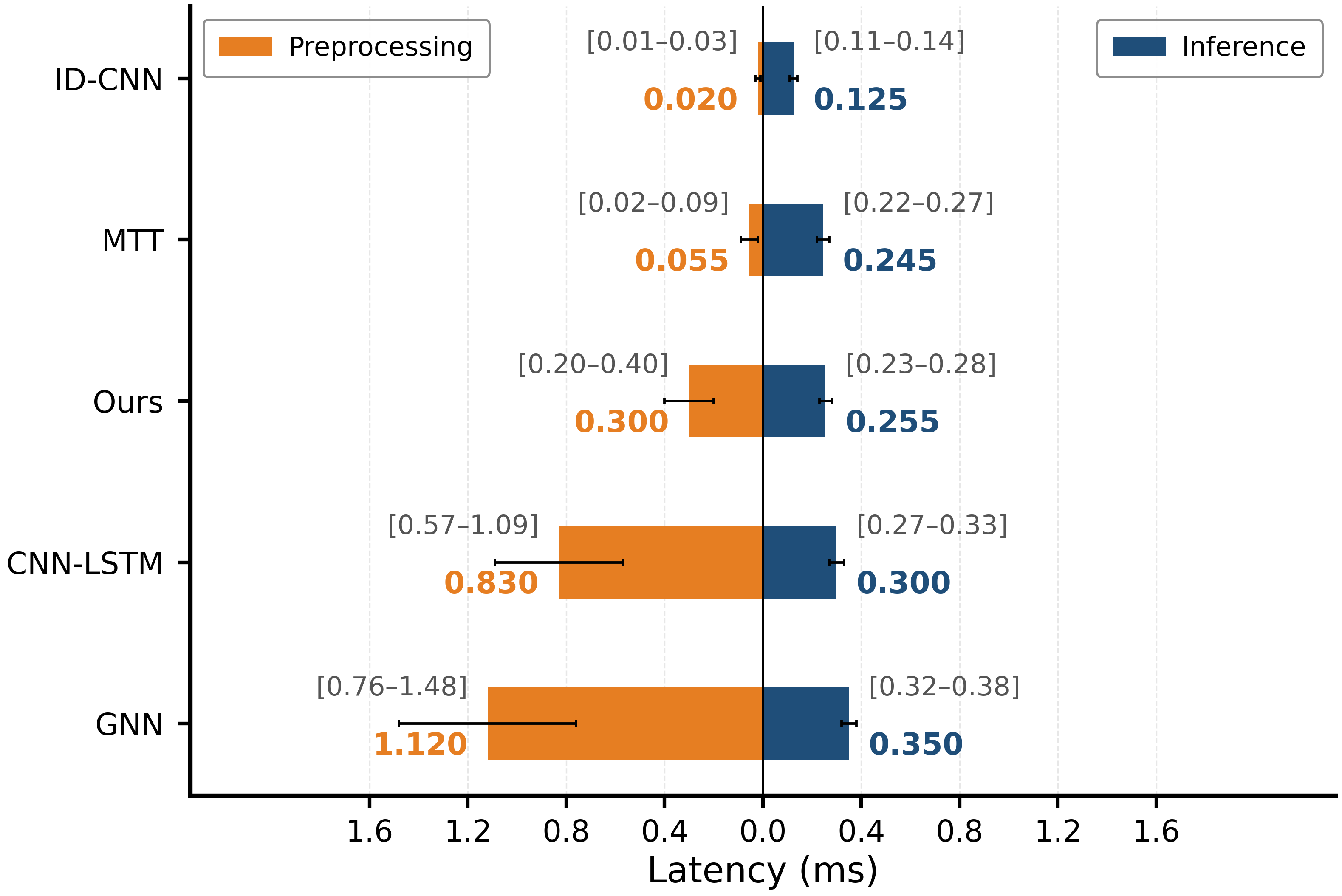} 
    \caption{End-to-end latency breakdown: inference latency and preprocessing latency with dynamic ranges across five classification methods.}
    \label{figc:single_fig}
\end{figure}

In urban rail train-to-ground edge deployment scenarios, computational overhead and end-to-end latency are critical determinants of engineering feasibility.  With only $0.18$~M parameters (raw‑traffic branch: $0.1304$~M, 73.7\%; statistical‑feature branch: $0.0376$~M, 21.2\%; remaining parameters come from channel‑attention module and classification head) and $23.8$~M FLOPs as shown in Fig.~\ref{figa:single_fig} , the proposed model is more lightweight than both MTT ($41.5$~M FLOPs) and 1D-CNN ($39.6$~M FLOPs). Fig.~\ref{figc:single_fig} breaks down the end-to-end latency of each model into preprocessing and inference stages. The proposed method incurs a higher preprocessing overhead (approximately 55\% of the total latency, on resource‑constrained edge hardware, this preprocessing step may consume a certain amount of CPU cycles and constitute a non‑negligible component of the overall system load.
) than 1D-CNN and MTT (approximately 20\%), reflecting the cost of constructing two parallel input representations. However, its forward inference latency ($0.23$--$0.28$~ms) is competitive with MTT ($0.22$--$0.27$~ms), and the end-to-end latency ($0.4$--$0.7$~ms)(Ranges are conservatively rounded outward (floor/ceil) to guarantee coverage of the measured interval.) remains well within the millisecond-level real-time constraint of CBTC.Thus, the model achieves near-perfect recall on critical services while maintaining competitive latency and the smallest parameter count among compared methods---a practical balance between classification accuracy and edge-deployment efficiency.



\subsection{Ablation Study on Module}
\label{subsec:ablation_modules}

\subsubsection{Experiment Design}
This section is dedicated to systematically validate the contributions of Transfer Learning, Dual-Channel Fusion, and Channel Attention (CA). We design five configurations: (1) Raw channel only (Raw); (2) Raw channel with transfer learning (Raw-TL); (3) Statistical channel only (Stats); (4) Dual-channel without channel attention (Dual CA-OFF); (5) the complete model with channel attention (Dual CA-ON). We additionally include an MLP baseline that takes the identical 16 statistical features as a flat vector input (3 fully-connected layers, $16\rightarrow256\rightarrow128\rightarrow5$, with ReLU and dropout), without the 2D reshaping step. This configuration isolates the contribution of the convolutional design.
All configurations in Table 4 are trained under the full Priority-Sensitive Loss to isolate the architectural contribution.

\begin{figure*}[!tp]
\centering
\includegraphics[trim=0 8 0 8mm, clip, width=\textwidth]{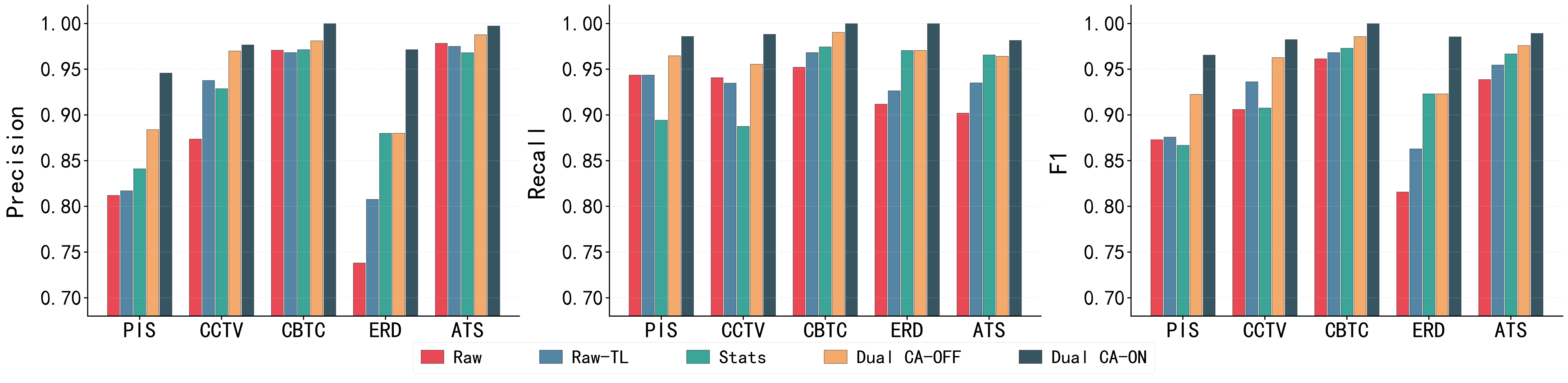}

\vspace{3pt} 
\begin{minipage}{0.31\textwidth}
\centering
(a) Per-Class Precision
\end{minipage}
\begin{minipage}{0.36\textwidth} 
\centering
(b) Per-Class Recall
\end{minipage}
\begin{minipage}{0.30\textwidth} 
\centering
(c) Per-Class F1
\end{minipage}

\caption{Per-class precision, recall, and F1-score across ablation configurations (Raw, Raw-TL,Stats, Dual CA-OFF, Dual CA-ON)}
\label{fig:ablation_per_class}
\end{figure*}

\begin{table*}[!tbp]
\centering
\footnotesize
\setlength{\tabcolsep}{3pt}
\caption{Ablation results: individual and combined contributions of the raw channel, statistical channel, and channel attention}
\label{tab:ablation}
\hspace*{-6mm}\begin{tabular}{cccccccc}
\toprule
\textbf{Configuration} & \textbf{Accuracy} & \textbf{Macro-Precision} & \textbf{W-Precision} & \textbf{Macro-Recall} & \textbf{W-Recall} & \textbf{Macro-F1} & \textbf{W-F1} \\
\midrule
MLP
& 91.49 ($\pm$ 0.15)
& 86.05 ($\pm$ 0.37)
& 86.63 ($\pm$ 0.47)
& 91.17 ($\pm$ 0.31)
& 92.67 ($\pm$ 0.31)
& 88.07 ($\pm$ 0.36)
& 89.00 ($\pm$ 0.40) \\
\cmidrule(lr){1-8}
Stats
& 94.51 ($\pm$ 0.11)
& 91.26 ($\pm$ 0.42)
& 91.92 ($\pm$ 0.65)
& 93.93 ($\pm$ 0.29)
& 95.57 ($\pm$ 0.31)
& 92.47 ($\pm$ 0.32)
& 93.61 ($\pm$ 0.45) \\
Raw
& 92.38 ($\pm$ 0.10)
& 88.05 ($\pm$ 0.47)
& 88.09 ($\pm$ 0.75)
& 92.93 ($\pm$ 0.35)
& 92.74 ($\pm$ 0.44)
& 90.22 ($\pm$ 0.37)
& 90.16 ($\pm$ 0.53) \\
Raw-TL
& 94.19 ($\pm$ 0.08)
& 90.68 ($\pm$ 0.41)
& 90.87 ($\pm$ 0.65)
& 94.43 ($\pm$ 0.22)
& 94.74 ($\pm$ 0.27)
& 92.41 ($\pm$ 0.31)
& 92.65 ($\pm$ 0.46) \\
Dual (CA-OFF)
& 96.85 ($\pm$ 0.08)
& 94.68 ($\pm$ 0.40)
& 94.87 ($\pm$ 0.65)
& 97.11 ($\pm$ 0.19)
& 97.57 ($\pm$ 0.23)
& 95.83 ($\pm$ 0.26)
& 96.15 ($\pm$ 0.39) \\
Dual (CA-ON)
& \textbf{98.74 ($\pm$ 0.05)}
& \textbf{97.82 ($\pm$ 0.21)}
& \textbf{97.93 ($\pm$ 0.31)}
& \textbf{99.00 ($\pm$ 0.17)}
& \textbf{99.24 ($\pm$ 0.23)}
& \textbf{98.40 ($\pm$ 0.14)}
& \textbf{98.57 ($\pm$ 0.21)} \\
\bottomrule
\end{tabular}\hspace*{-6mm}
\end{table*}

\subsubsection{Results Analysis}
Table~\ref{tab:ablation} reports the overall accuracy and weighted metrics for the five model configurations. Single-channel architectures exhibit clear bottlenecks. The raw channel alone yields the lowest accuracy at $92.38 (\pm0.10)\%$. Introducing transfer learning (Raw-TL) improves accuracy by $1.81$~pp to $94.19 (\pm0.08)\%$, with the gain primarily concentrated on control services, demonstrating that source-domain knowledge provides effective priors for structurally stable traffic. The statistical channel achieves slightly higher accuracy than Raw-TL ($94.51 \pm0.11\%$) but underperforms on multimedia services (see Fig.~\ref{fig:ablation_per_class}). For comparison, an MLP processing the same 16 statistical features as a flat vector achieves only $91.49  (\pm0.15)\%$, given the same 16 statistical features, convolutional processing outperforms treating the features as independent inputs. Dual-channel fusion (Dual CA-OFF) substantially boosts accuracy to $96.85 (\pm0.08)\%$, confirming the complementarity between raw textures and statistical features. Incorporating channel attention (Dual CA-ON) further pushes the model to $98.74 (\pm0.05)\%$, with the smallest standard deviation among all configurations, confirming that dynamic weighting effectively stabilizes feature fusion and suppresses noise.

Fig.~\ref{fig:ablation_per_class} presents the per-class Precision, Recall, and F1-score across configurations, clearly demonstrating modality complementarity and the critical role of attention. Specifically, the Raw channel achieves significantly higher recall than the Stats channel on multimedia services (PIS, CCTV), as multimedia traffic contains rich local spatial textures in raw byte sequences. In contrast, the Stats channel performs better on periodic control services (ERD, ATS) because it effectively captures macro-level patterns such as inter-arrival time and burstiness. Transfer learning primarily improves the Raw channel's generalization on control services but does not fully close the gap with the Stats channel. Simple dual-channel concatenation (Dual CA-OFF) already combines the strengths of both modalities, leading to substantial recall improvements across all classes, yet some classes (e.g., ERD, CCTV) remain suboptimal. After incorporating attention, the model adaptively suppresses noise from redundant modalities and amplifies discriminative features, further compensating for the weaknesses of each single channel: recall on control services (where Raw was weak) and on multimedia services (where Stats was weak) are both significantly enhanced. Consequently, recall exceeds 98\% for all five business categories, achieving near-zero missed detection for safety-critical services (CBTC, ERD). The ablation study conclusively validates the synergistic gains of transfer learning, dual-channel fusion, and channel attention, demonstrating the proposed architecture's superior robustness and accuracy in complex urban rail traffic classification.

\subsection{Ablation Study on Loss Function}
\label{subsec:loss_ablation}

This section is dedicated to validating the effectiveness of the dual-weight design in the proposed PSL.

\subsubsection{Experiment Design}
We compare four loss function configurations: (1) Standard Cross-Entropy; (2) Prior only (priority weighting without balance); (3) Balance only (class balance without priority rescaling); and (4) Ours (full PSL with both business priority weight $\omega_c^{\mathrm{prior}}$ and class balance weight $\omega_c^{\mathrm{bal}}$). All configurations share the identical Dual CA-ON architecture and hyperparameters.All loss configurations, including the cross-entropy baseline, were trained and evaluated under the identical class-wise temporal 6:2:2 split with the same five random seeds.

\begin{table*}[!tbp]
\centering
\footnotesize
\setlength{\tabcolsep}{3pt}
\caption{Ablation of loss function design: cross‑entropy vs. balance‑only, priority‑only, and the proposed PSL}
\label{tab:loss_ablation}
\hspace*{-6mm}\begin{tabular}{cccccccc}
\toprule
\textbf{Loss Function} & \textbf{Accuracy} & \textbf{Macro‑Precision} & \textbf{W‑Precision} & \textbf{Macro‑Recall} & \textbf{W‑Recall} & \textbf{Macro‑F1} & \textbf{W‑F1} \\
\midrule
Cross‑Entropy
& 93.61 ($\pm$ 0.16)
& 89.89 ($\pm$ 0.58)
& 90.23 ($\pm$ 0.84)
& 92.92 ($\pm$ 0.49)
& 93.28 ($\pm$ 0.59)
& 91.29 ($\pm$ 0.49)
& 91.65 ($\pm$ 0.66) \\
Balance‑only
& 95.67 ($\pm$ 0.09)
& 92.74 ($\pm$ 0.36)
& 92.89 ($\pm$ 0.60)
& 96.15 ($\pm$ 0.18)
& 96.63 ($\pm$ 0.16)
& 94.33 ($\pm$ 0.27)
& 94.64 ($\pm$ 0.36) \\
Prior‑only
& 96.02 ($\pm$ 0.08)
& 93.33 ($\pm$ 0.35)
& 93.58 ($\pm$ 0.65)
& 96.58 ($\pm$ 0.28)
& 97.54 ($\pm$ 0.33)
& 94.85 ($\pm$ 0.29)
& 95.44 ($\pm$ 0.46) \\
Ours
& \textbf{98.74 ($\pm$ 0.05)}
& \textbf{97.82 ($\pm$ 0.21)}
& \textbf{97.93 ($\pm$ 0.31)}
& \textbf{99.00 ($\pm$ 0.17)}
& \textbf{99.24 ($\pm$ 0.23)}
& \textbf{98.40 ($\pm$ 0.14)}
& \textbf{98.57 ($\pm$ 0.21)} \\
\bottomrule
\end{tabular}\hspace*{-6mm}
\end{table*}

\subsubsection{Results Analysis}
\begin{figure}[!tbp]
\centering
\begin{subfigure}[b]{0.48\columnwidth}
    \centering
    \includegraphics[width=\linewidth]{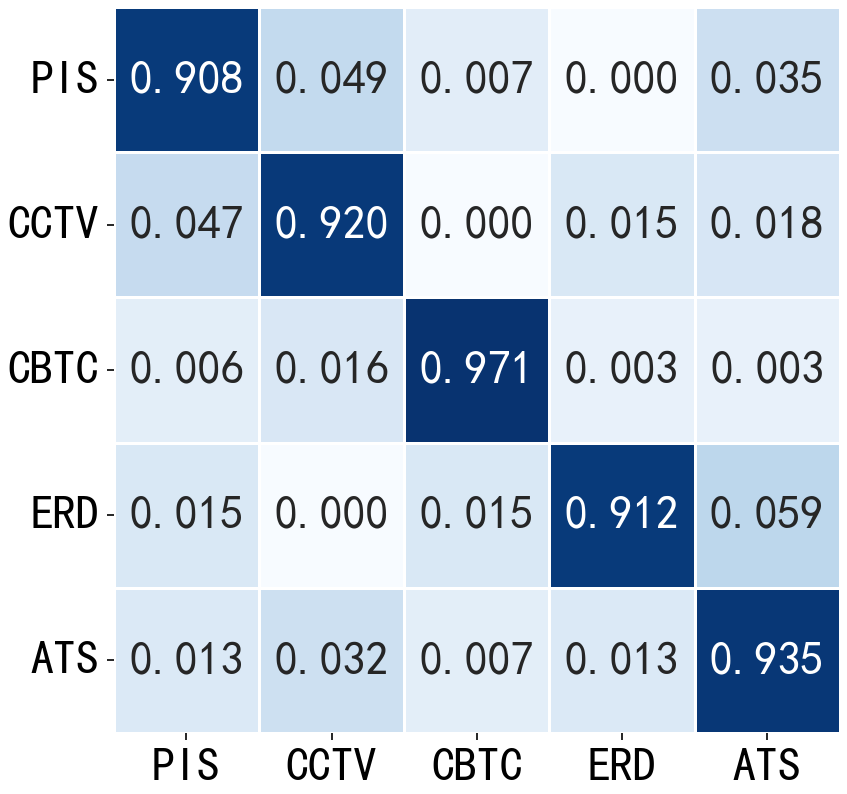}
    \caption{Cross-entropy}
\end{subfigure}
\hfill 
\begin{subfigure}[b]{0.48\columnwidth}
    \centering
    \includegraphics[width=\linewidth]{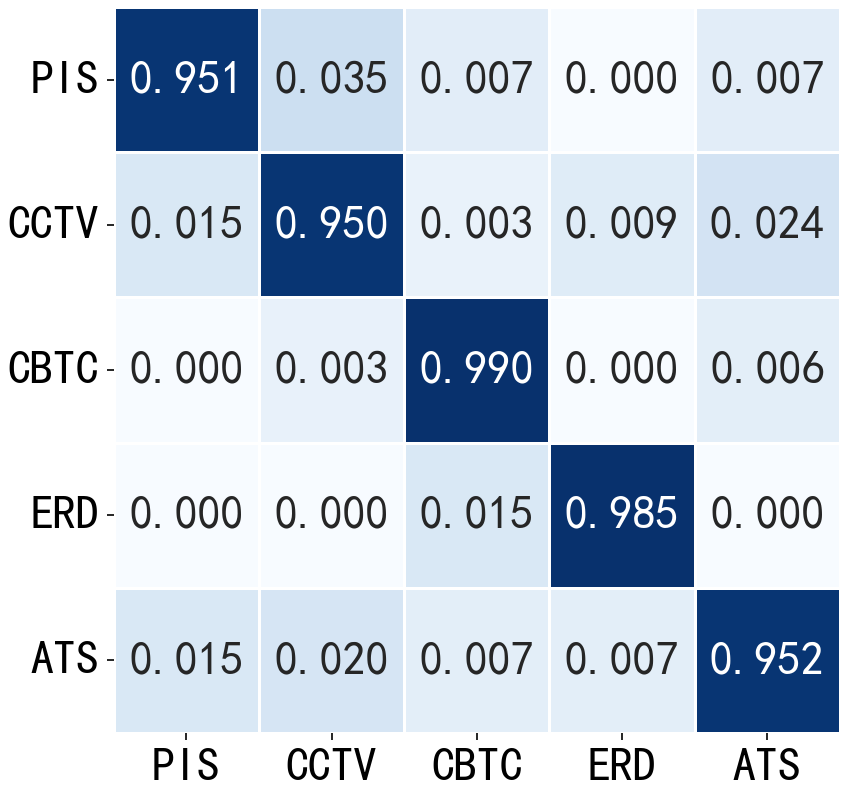}
    \caption{Prior Weight Only}
\end{subfigure}

\vspace{6pt} 

\begin{subfigure}[b]{0.48\columnwidth}
    \centering
    \includegraphics[width=\linewidth]{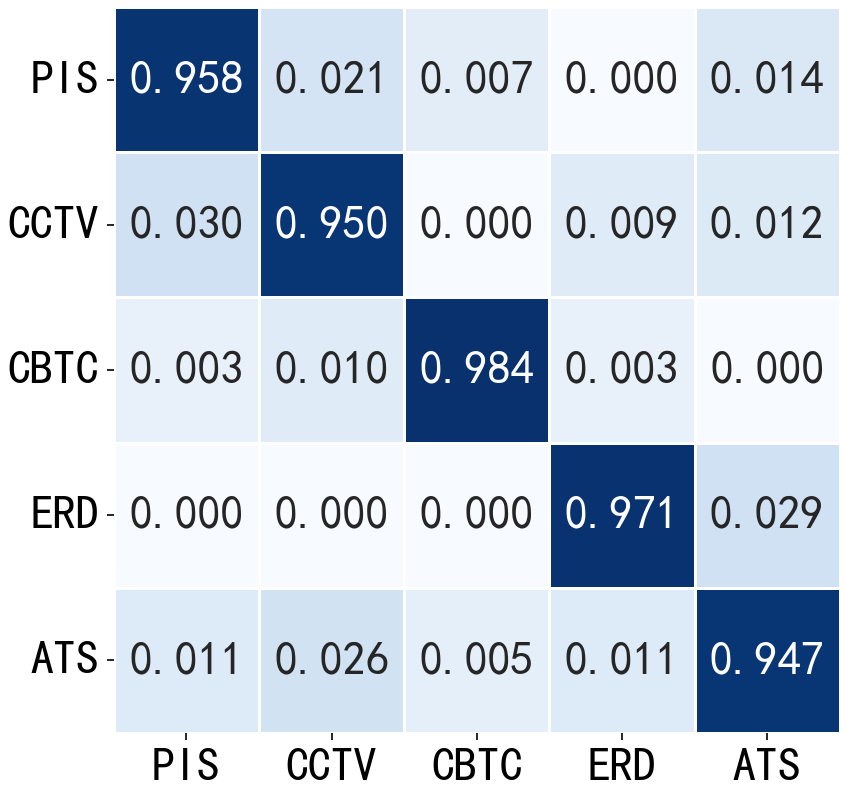}
    \caption{Balance Weight Only}
\end{subfigure}
\hfill
\begin{subfigure}[b]{0.48\columnwidth}
    \centering
    \includegraphics[width=\linewidth]{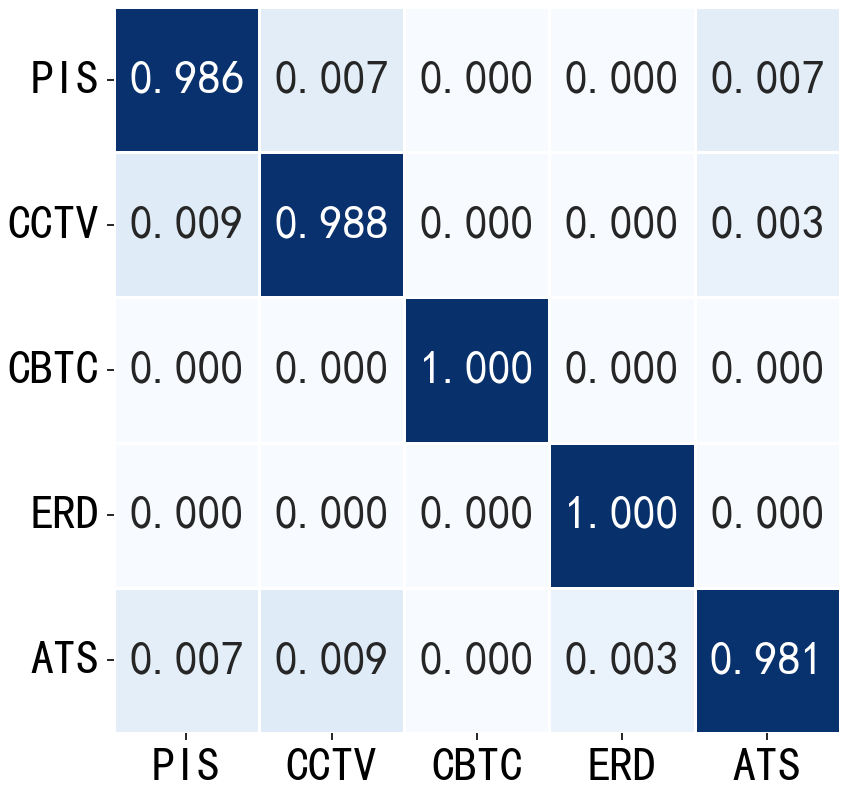}
    \caption{Ours (Dual weights)}
\end{subfigure}

\caption{Confusion matrices comparison of different loss function configurations.}
\label{fig:loss_ablation_cm}
\end{figure}

Table~\ref{tab:loss_ablation} summarizes the overall accuracy and weighted metrics for the four loss function configurations. Additionally, to intuitively illustrate the fine-grained classification performance across different service types, the confusion matrix on the test set is presented in Fig.~\ref{fig:loss_ablation_cm}.

It clearly visualizes the per-class prediction accuracy and misclassification patterns, further validating the effectiveness of the dual-channel fusion architecture under complex multi-service scenarios.
The full PSL achieves the best performance across all indicators: accuracy of $98.74 (\pm0.05)\%$ and weighted recall of $99.24 (\pm0.23)\%$. With Prior only, safety-critical services (CBTC and ERD) are well preserved, but recall on low-priority services degrades, yielding a weighted recall of $97.54\%$. Balance only improves minority-class recall at some cost to the majority class, resulting in a weighted recall of $96.63\%$. The unweighted cross-entropy performs worst across all metrics; notably, recall on the safety-critical ERD class drops to $90.59\%$, confirming that without explicit weighting, the gradient is dominated by the most populous class (ATS). These results demonstrate that the two weighting factors serve complementary roles: $\omega_{bal}$ counters the raw sample-size imbalance, while $\omega_{prior}$ directs the optimizer toward operationally critical services. Neither acting alone achieves both objectives simultaneously.

The off-diagonal structure of the confusion matrices in Fig.~\ref{fig:loss_ablation_cm} further clarifies the complementary roles of the two weights. Under cross-entropy, the dominant pattern is a systematic drift of minority-class predictions toward the majority class ATS, regardless of service priority, PIS, CCTV, and ERD are all absorbed into ATS at non-negligible rates. Under balance-only, this majority absorption is substantially reduced, but residual leakage from the smallest classes (particularly ERD) toward ATS persists, because inverse-frequency weighting at the loss level cannot fully compensate for an order-of-magnitude sample-size disparity. Under prior-only, the direction of confusion reverses: ATS samples themselves begin to be misclassified as the highest-priority service ERD, indicating that priority weighting without a balance term overcorrects the decision boundary, pulling majority-class samples toward the safety-critical prediction slots. Only the full PSL, which couples $\omega_{prior}$ with $\omega_{bal}$, avoids both failure modes: majority absorption is eliminated and no spurious high-priority over-prediction occurs.


\section{Conclusion}
\label{sec:conclusion}

This paper proposed a priority-aware dual-channel feature fusion framework for classifying encrypted urban rail service traffic. The raw channel, enhanced by ISCX pre-training, extracts protocol-level byte textures; the statistical channel captures flow-level temporal and distributional patterns; channel attention adaptively reweights these two modalities on a per-sample basis; and the PSL directs the optimization toward the operational priority structure, jointly weighting business criticality and class balance. Together, these components form an end-to-end pipeline that achieves robust identification of safety-critical services on resource-constrained edge hardware.

Under a temporal split with five independent seeds, the method achieves $98.74 (\pm0.05)\%$ accuracy and $99.24 (\pm0.23)\%$ weighted recall, with CBTC and ERD recall reaching $99.94\%$ and $99.41\%$. The model requires only $0.18$M parameters and $0.4$--$0.7$~ms end-to-end latency. Ablation studies confirm that the raw and statistical channels contribute complementary information and that channel attention further closes residual performance gaps on both multimedia and control services.

Despite these promising results, several limitations should be acknowledged.
(1) The statistical confidence of the metrics for the minority class is inherently limited. Under the temporal split, the ERD test set contains only 68 flows, so its recall is quantized in steps of approximately $1.47\%$ ($1/68$); the reported $99.41\%$ recall therefore corresponds to zero or one misclassified sample per run. By the rule of three, the 95\% confidence upper bound on the missed-detection rate is approximately $4\%$ for ERD and $1\%$ for CBTC, which should be taken into account when interpreting the near-perfect recall of the safety-critical services. 
(2)The five random seeds control weight initialization, data shuffling, and dropout on a single fixed temporal split; the reported standard deviations thus reflect training stochasticity rather than sampling uncertainty, since the same 68 ERD test flows are evaluated in all five runs. Second, the dataset was collected from a single operational line over a limited period and is proprietary, so the generalizability of the proposed method to other lines, vendors, or operational environments requires further validation. 
(3)Although the per-class temporal split prevents leakage of temporally adjacent flows between the training and test sets, residual correlation among flows within the same session may still exist.

\section{Future Work}
\label{sec:futurework}

Future work will advance this framework toward production deployment along several directions. Model compression via quantization and structured pruning will reduce inference latency to sub-millisecond levels on low-power on-board processors, enabling integration with existing vehicle-edge hardware. Self-supervised pretraining on unlabeled urban rail traffic, coupled with validation across multiple lines and extended operational periods, will improve generalization and reduce reliance on manually labeled data. Robustness under complex dynamic network conditions, including channel fading, handover-induced packet loss, and bursty interference typical of train-to-ground wireless links, warrants dedicated study. On the application side, the classification output can directly serve as the perception front-end for downstream tasks such as QoS-aware resource scheduling, anomaly detection for equipment fault conditions, and open-set identification of unknown service types, forming a practical bridge from traffic awareness to intelligent network operation and maintenance.

\backmatter

\section*{Acknowledgements}

The authors would like to thank the anonymous reviewers for their valuable comments and the Institute of Computing Technologies, China Academy of Railway Sciences Corporation Limited.

\section*{Author Contributions}

Xinpeng Liu: Conceptualization, Methodology, Software, Investigation, Formal Analysis, Data Curation, Original Draft, Review \& Editing, Visualization;
Junhui Zhao: Conceptualization,Review \& Supervision;
Zhengyuan Wu: Conceptualization, Methodology;
Huaicheng Li: Review \& Supervision.

\section*{Funding}

This work was supported by National Engineering Research Center of System Technology for High Speed Railway and Urban Rail Transit, under Grant No.~2024YJ255.

\section*{Declarations}

\subsection*{Conflict of Interest}

The authors declare that they have no known competing financial interests or personal relationships that could have appeared to influence the work reported in this paper.

\subsection*{Data Availability}

The VPN-NonVPN Dataset (ISCXVPN2016) is provided by the Canadian Institute for Cybersecurity (CIC), University of New Brunswick (UNB), and can be accessed at: \url{http://www.unb.ca/cic/datasets/vpn.html}.
The urban rail transit traffic dataset used in this study is provided by the Institute of Computing Technologies, China Academy of Railway Sciences Corporation Limited. Due to project confidentiality and privacy-related restrictions, the raw data cannot be made publicly available at the present stage.

\bibliography{cas-refs}

\end{document}